\documentclass[twocolumn,prl,amsmath,amssymb,showpacs,superscriptaddress,floatfix]{revtex4}
\usepackage{graphicx}
\usepackage{bm}
\usepackage{lineno,hyperref}
\usepackage{epsf}
\usepackage{xcolor}
\usepackage{epstopdf}
\begin{document}
\title{Crystal structures of a core-softened system confined in a narrow slit pore}

\author{Yu. D. Fomin \footnote{Corresponding author: fomin314@mail.ru}}
\affiliation{Vereshchagin Institute of High Pressure Physics, Russian Academy of Sciences,
Kaluzhskoe shosse, 14, Troitsk, Moscow, 108840 Russia}
\affiliation{Moscow Institute of Physics and Technology (National Research University), 9 Institutskiy Lane, Dolgoprudny, Moscow region, 141701, Russia}

\author{E. N. Tsiok}
\affiliation{Vereshchagin Institute of High Pressure Physics, Russian Academy of Sciences,
Kaluzhskoe shosse, 14, Troitsk, Moscow, 108840 Russia}

\author{V. N. Ryzhov}
\affiliation{Vereshchagin Institute of High Pressure Physics, Russian Academy of Sciences,
Kaluzhskoe shosse, 14, Troitsk, Moscow, 108840 Russia}

\date{\today}

\begin{abstract}

We investigate a behavior of a core-softened system in a thin slit pore (the width of the pore is equil to three diameters of the particles). In previous studies it was shown that strongly confined systems form crystalline phases which consist of several triangular or square layers. These phases can be also considered as cuts of FCC or HCP structures. We show that the behavior of the core-softened system is more complex. We observe also a quasicrystalline phase. Moreover, the phase with two triangular layers appears at lower densities than the one with two square layers which is in contrast to the behavior of the systems studied before. These results demonstrate that the phase behavior of strongly confined systems can be even more complex than it was supposed before.

\end{abstract}

\pacs{61.20.Gy, 61.20.Ne, 64.60.Kw}

\maketitle

\section{Introduction}

It is well known that confinement strongly affects many properties of a substance, including thermodynamic properties (for instance, melting temperature), structure of the system (density modulations) and dynamic properties (for instance, diffusion constant) \cite{rice-review,review-2}. Moreover, the effect of confinement depends on the shape of the pore and the interaction between the confined particles and the walls. All of this makes the properties of confined systems to be very reach.

Among the most important properties of confined systems is their ability to form some structures which are not observed in the corresponding bulk systems. For instance, many unusual phases were reported for water confined in nanotubes \cite{wat-nanotube,wat-nanotube-1}. Confined metals also can crystallize into unusual structures. For instance, in Ref. \cite{iron} a two-dimensional (2d) square crystal of iron confined in a graphene slit pore was observed.

Most of studies of crystallization of confined liquids were performed with colloidal system. Usually colloidal systems can be described by relatively simple interactions \cite{likos-interaction}. In many cases the interaction between particles can be described as hard sphere interaction or Yukawa potential. The walls of the pore can be treated as structureless and the particle-wall interaction is effectively determined by a simple pair potential, such as Lennard-Jones (LJ) 9-3 potential or Steel potential, which depends on the distance from the particle to the wall only. It is much harder to perform experiments with atomic systems. However, the Yukawa interaction can be considered as a very crude model of metal, therefore some conclusions obtained for colloidal systems can be transfered to the metallic systems too.

The general form of the structure of colloidal system in strong confinement is studied in a great number of works, both experimental and theoretical \cite{pansu,pansu1,pansu2,winkle,video,buck,buck1,lj1,lj2,lj3,lj4,lj5,oguz1,oguz2,oguz3,conf1,conf2,conf3}. Most of the works describe the structure of the system confined in a slit pore as a number of two-dimensional layers with triangular or square symmetry. Going from thin pore to the thiker ones the sequence of phases is 1$\Delta$ -
2$\fbox{$\phantom{5}$}$ - 3$\Delta$ - 3$\fbox{$\phantom{5}$}$, etc., where the number is the number of layers in the system,
$\Delta$ denotes the triangular structure within each layer and $\fbox{$\phantom{5}$}$ corresponds to the square structure of the layers. Some of the studies report also rhombic phase \cite{oguz2},  hcp-like phases and prism phases \cite{prism}. The prism phase can be considered as FCC crystal stacks with a periodic pattern of stacking faults \cite{prism}. One more phase appearing in the system is so called buckled phase \cite{buck,buck1}. In this phase the particles are located in a zigzag order.

The description of the structure of a system confined in a slit pore as a multilayered 2d crystal is valid up to the width of the pore about seven diameters of the particles \cite{pansu1}.

In our recent work \cite{ljconf} the approach to the description of the structure of a confined system as a set of 2d layers was critisized. Basing on the bond orientational order parameters \cite{bondorder} it was shown that all multilayered systems can be reduced to a cut of HCP or FCC crystal which can be turned to the walls by different surfaces. The triangular 2d crystals corresponds to the (111) face of FCC phase, while square crystal can be considered as (100) face of FCC. Although the idea that multilayered crystals can be considered as cuts of FCC or HCP appeared long time ago \cite{pansu1}, it was not accepted for two reasong. First, only a discrete set of lattice constants can be consistent with the lattice spacing both in the plane of the walls and in the perpendicular direction. Second, a buckled phase was considered as a special phase which does not have an anolog in 3d space. In our work it was shown that the first problem is overcome by intrinsic defects of spontaneously freezing confined crystal. This is perfectly consistent with the prism phase mentioned above: the prism phase is a stack of FCC clusters separated by staking faults \cite{prism}. The second item is solved by noticing that the buckled phase gives the bond orientational order consistent with HCP phase and therefore the buckled phase is a special orientation of HCP crystal. This approach allows to describe the whole phase diagram of the crystals in a slit pore on the same footing: the structure is either HCP or FCC with intrinsic defects. All phases observed in a slit pore before (n$\Delta$, n$\fbox{$\phantom{5}$}$, HCP-like, buckled phase and prism phase) are consistent with this description. The rhombic phase is most probably a distorsion of HCP or FCC crystal.

Most of the previous studies both experimental and theoretical considered the systems with relatively simple phase diagrams. In the case of theoretical treatments it was Hard Spheres (see, i.e., Ref. \cite{prism}), Lennard-Jones system  (i.e., \cite{lj1,lj2,lj3,lj4,lj5,ljconf}) or Yukawa particles (i.e., \cite{oguz1,oguz2,oguz3}). In the case of experimental studies mostly colloidal suspensions were studied. The interaction of the colloids in these suspensions can be described by the same theoretical models. A set of works on a system with more complex interaction was reported in Refs. \cite{rice1,rice2,rice3}. The interaction potential of this system belongs to a family of core-softened systems, which can demonstrate very complex phase diagrams. However, the general trends in the system were qualitatively similar to the ones of the simple systems. Moreover, the phase diagram of both 2d and 3d system studied in these works in unknown, which makes the analysis of the confined system harder.

Another core-softened model was introduced in our recent works \cite{3d3}. This system is called Repulsive Shoulder System (RSS). It is characterized by the interaction potential of the form:

\begin{equation}
  U(r)/\varepsilon= \left( \frac{d}{r} \right)^n+0.5 \left( 1 - tanh(k(r- \sigma))
  \right),
\end{equation}
where $n=14$, $k=10$. The parameter $\sigma$ determines the
 width of the repulsive shoulder of the potential. In the present work it is set to $\sigma/d=1.35$. The parameters $\varepsilon$ and $d$ set the energy and length scales respectively. Below we express all quantities in the dimensionless units based one these parameters.

The phase diagrams of RSS in 2d \cite{2d1,2d2,2d3,2d4,2d5,2d6,2d7}, 3d \cite{3d3,3d1,3d2,3d4,3d5} and very strong confinement (only one layer of the particles in a slit pore) \cite{135-conf} are reported in our recent papers. It was shown that this system demonstrate very complex phase behavior. In 3d it has numerous crystal phases with different symmetry. Moreover, it demonstrates a glass transition \cite{3d3,rysch} which is rather unusual for a monatomic system. The phase diagram in 2d demonstrates some unusual features, such as square crystal and dodecagonal quasicrystal \cite{2d7}. The behavior of the system in very strong confinement (only one layer of the particles inside a slit pore) is qualitatively similar to the pure 2d one \cite{135-conf}.

In the present paper we extend study of Ref. \cite{135-conf} to the case of larger pores, i.e. to the system with two and three layers of particles in a slit pore. We demonstrate that the behavior of this system is more complex than the one of the hard sphere and Yukawa systems. In particular, the multilayer approach cannot be used to describe the structure of this system even in rather narrow pore with two or three layers of the particles. Taking into account that the effective potential of many liquid metals can be represented as a core-softened one, it leads us to the conclusion that the behavior of liquid metals in strong confinement can be much more complex than the one of colloids between glass plates.

\section{System and Methods}

In this work we study the RSS system in a slit pore by means of molecular dynamics simulation in canonical ensemble (constant number of particles N, volume V and temperature T). A system of 32000 particles is studied. The box has a square shape in X-Y plane (the plane of the walls). Periodic boundary conditions are used in this plane. The width of the pore is $H=3.0$. No periodic boundaries in Z direction are used. The density of the system is defined via a simple geometrical definition: $\rho=\frac{N}{L^2 H}$, where $L=L_x=L_y$ is the size of the system in X and Y directions. The densities from $\rho=0.3$ to $1.5$ with step $\Delta \rho=0.1$ were simulated. The initial configurations at given density were obtained by simulating the confined system at high temperature $T=10.0$. Then the system was quanched to low temperature $T=0.1$. All results presented in this paper correspond to this low temperature.

The interaction of the particles with the walls is modeled by LJ-9-3 potential:

\begin{equation} \label{wall}
U_{fw}=\varepsilon \left( \frac{2}{15} \left( \frac{\sigma}{z}
\right)^9  - \left( \frac{\sigma}{z} \right)^3 \right).
\end{equation}
The parameters of the potential are $\varepsilon=1$ , $\sigma=1$, $r_c=2.5$.

Firstly, the system was equilibrated for $1 \cdot 10^7$ steps. Then more $5 \cdot 10^7$ steps were made for calculation of the properties. The time step is set to $dt=0.001$.

We calculate the distribution of density along z-axis. At some densities the system splits in two or three well-defined layers. In this case we characterize the structure within each layer by a two-dimensional radial distribution function (RDFs). We also calculate the diffraction pattern if the system as $S(\bf{k})= < \frac{1}{N} \left( \sum_i^Ncos(\bf{kr}_i)\right)^2+\left( \sum_i^Nsin(\bf{kr}_i)\right)^2>$, where the wave-vector has the form ${\bf k}=(k_x,k_y,0)$, i.e. the intensities in X-Y plane are studied.

All simulations were performed using LAMMPS simulation package \cite{lammps}.

\section{Results and Discussion}

In this section we describe the results for different densities. The titles of the subsections correspond to the values of average density $\rho= \frac{N}{L^2H}$.

\subsection{Two layers}

\begin{figure}

\includegraphics[width=8cm, height=6cm]{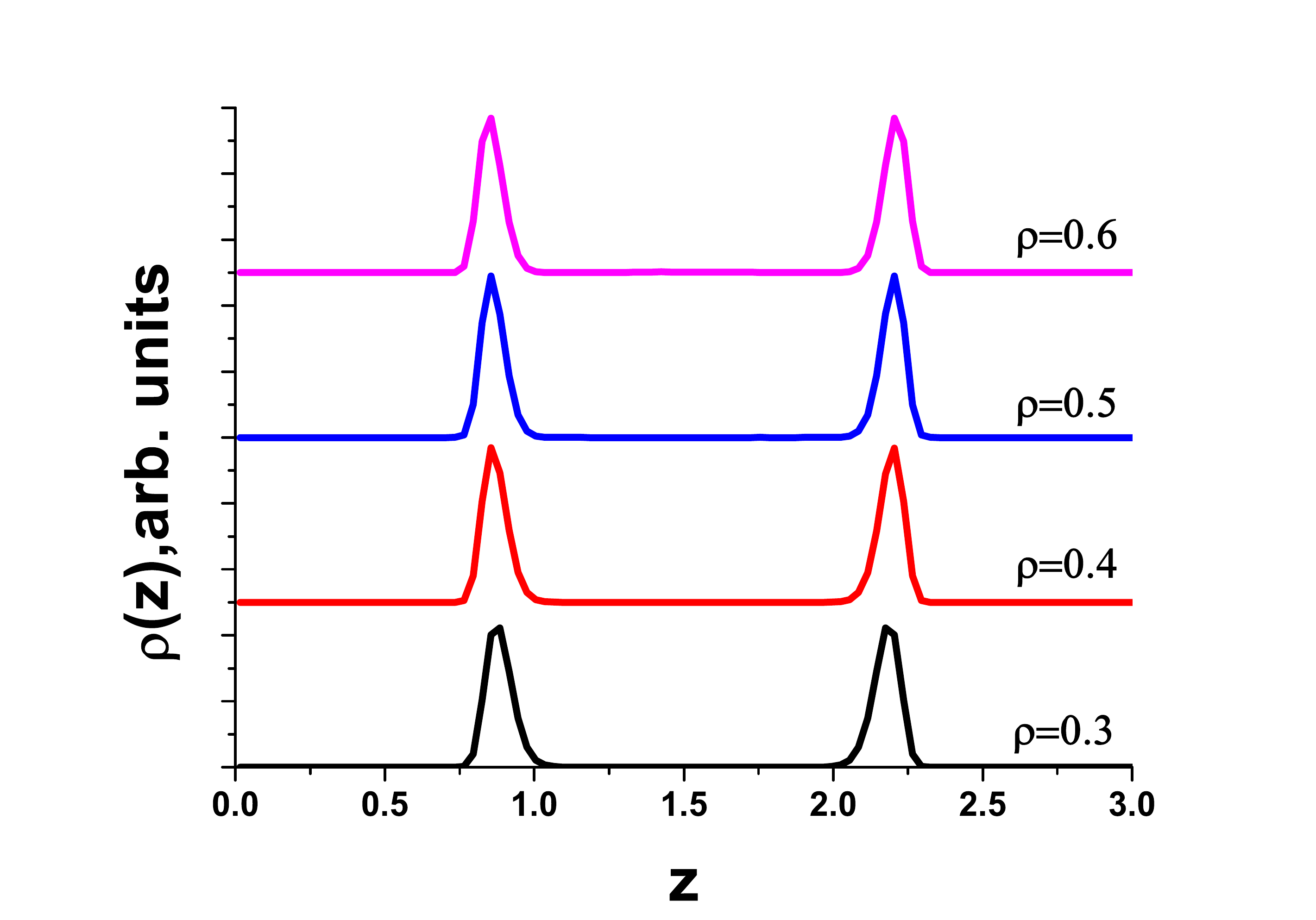}%

\caption{\label{z0306} Distribution of density along z axis for the average densities from 0.3 to 0.6. The curves are shifted with respect to each other to make the figure more clear.}
\end{figure}

At the densities from $\rho=0.3$ to $\rho=0.6$ the system splits into two layers. The distribution of the density along z axis is shown in Fig. \ref{z0306}.

\begin{figure}

\includegraphics[width=8cm, height=6cm]{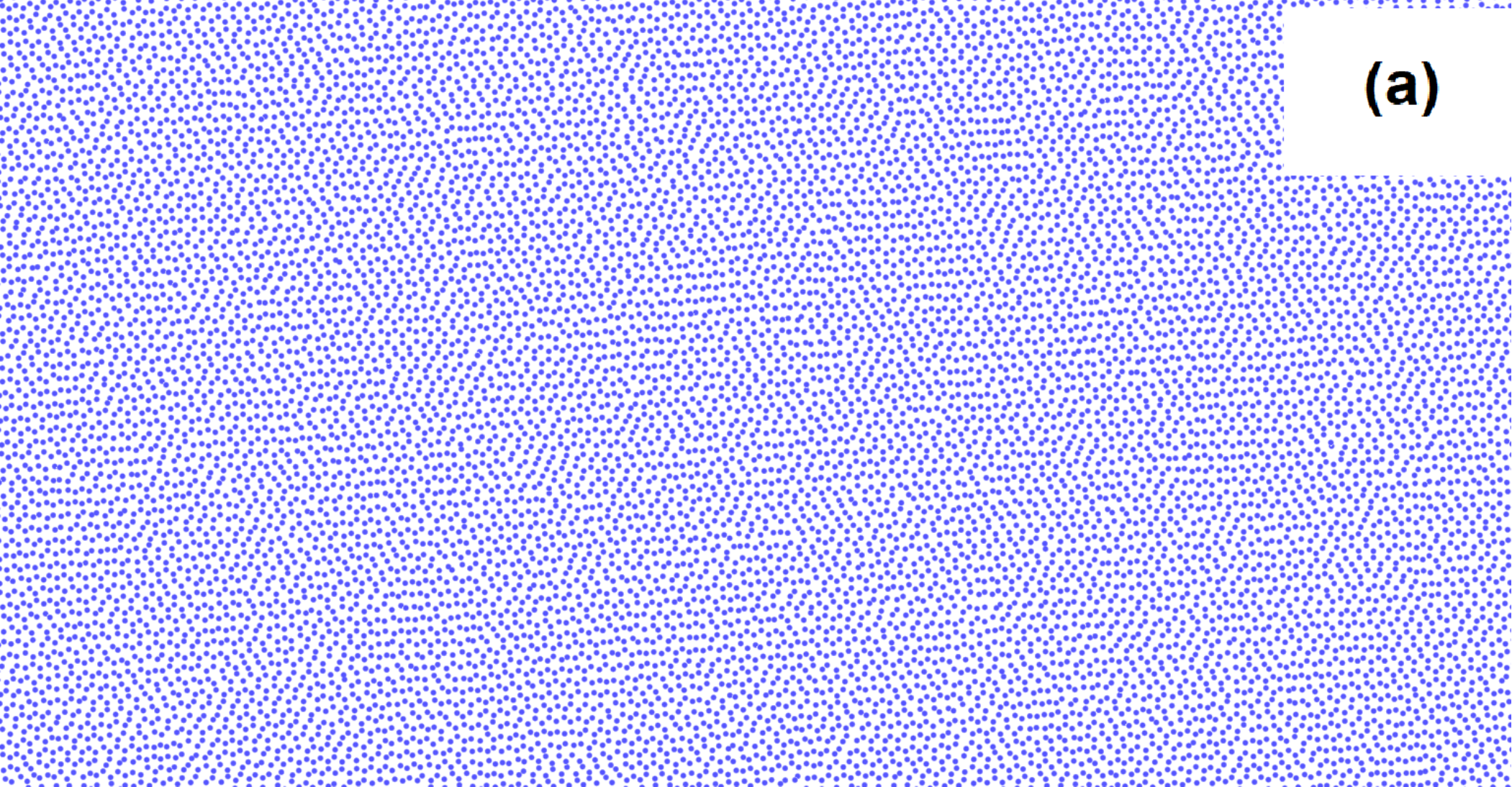}%

\includegraphics[width=8cm, height=6cm]{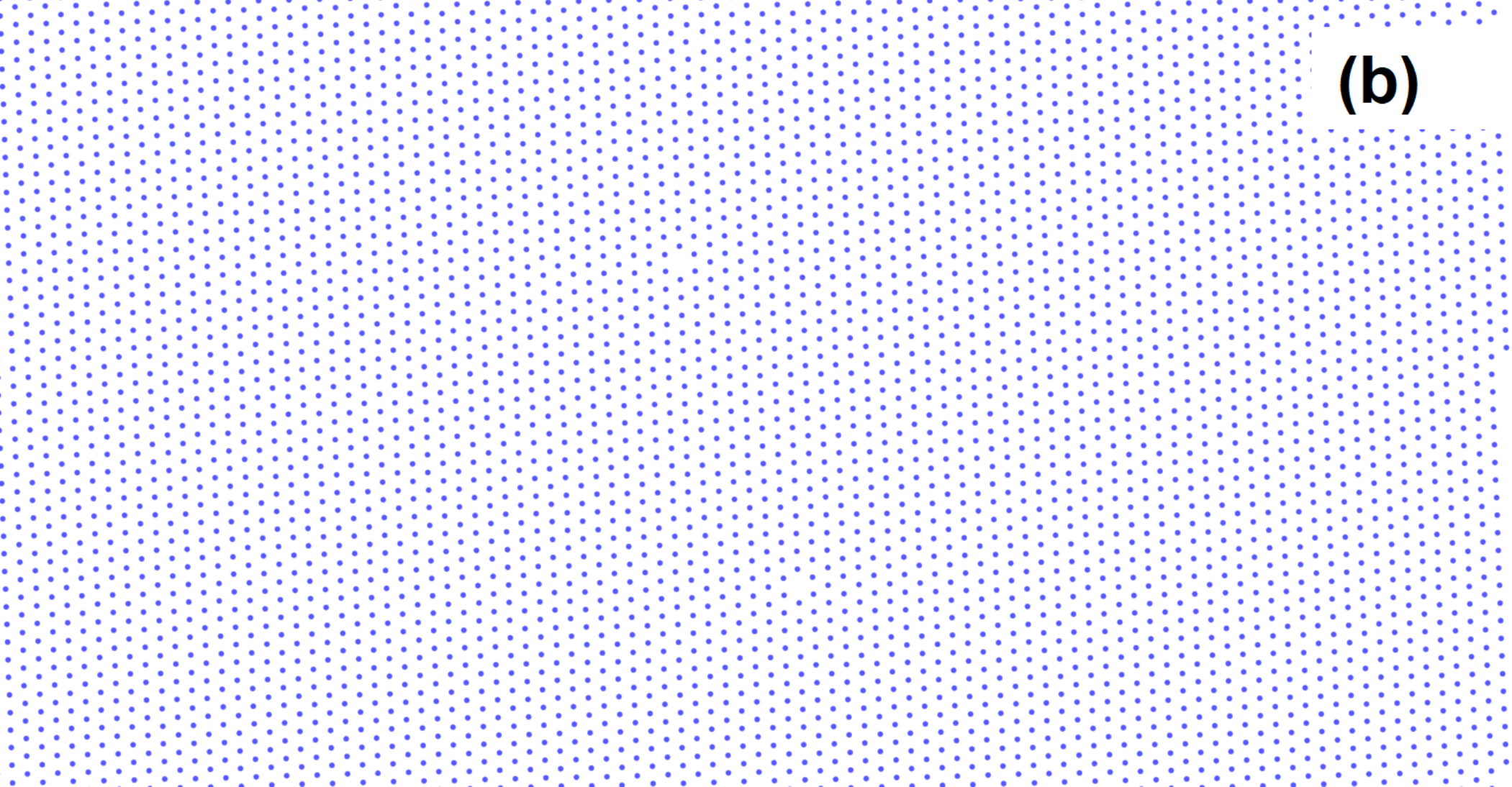}%

\includegraphics[width=8cm, height=8cm]{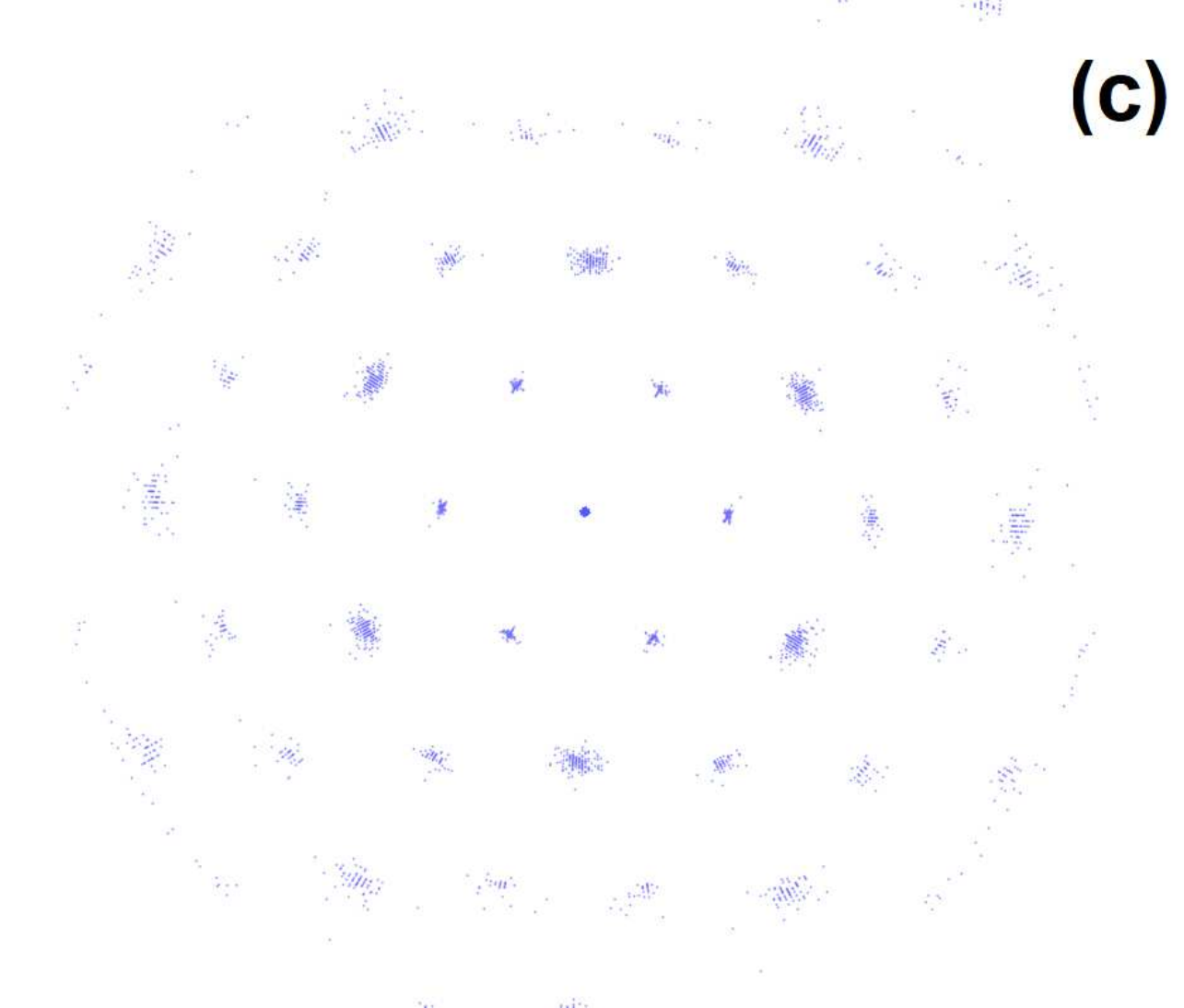}%

\caption{\label{r03} (a) A snapshot of the system at average density 0.3. (b) A snapshot of the first layer of the same system. (c) Diffraction pattern of the same system.}
\end{figure}

Fig. \ref{r03} (a) shows the snapshot of the system at $\rho=0.3$. Panel (b) of the same figure demonstrates a snapshot of the first layer, and panel (c) - the diffraction pattern of the whole system. Although it is not clearly seen from the snapshot, the system demonstrate HCP-like structure. It is confirmed by the diffraction pattern of the system. The structure of each layer is almost ideal triangular 2d crystal.

At the density $\rho=0.4$ the behavior of the system is qualitatively similar to the one at $\rho=0.3$.

\begin{figure}

\includegraphics[width=8cm, height=6cm]{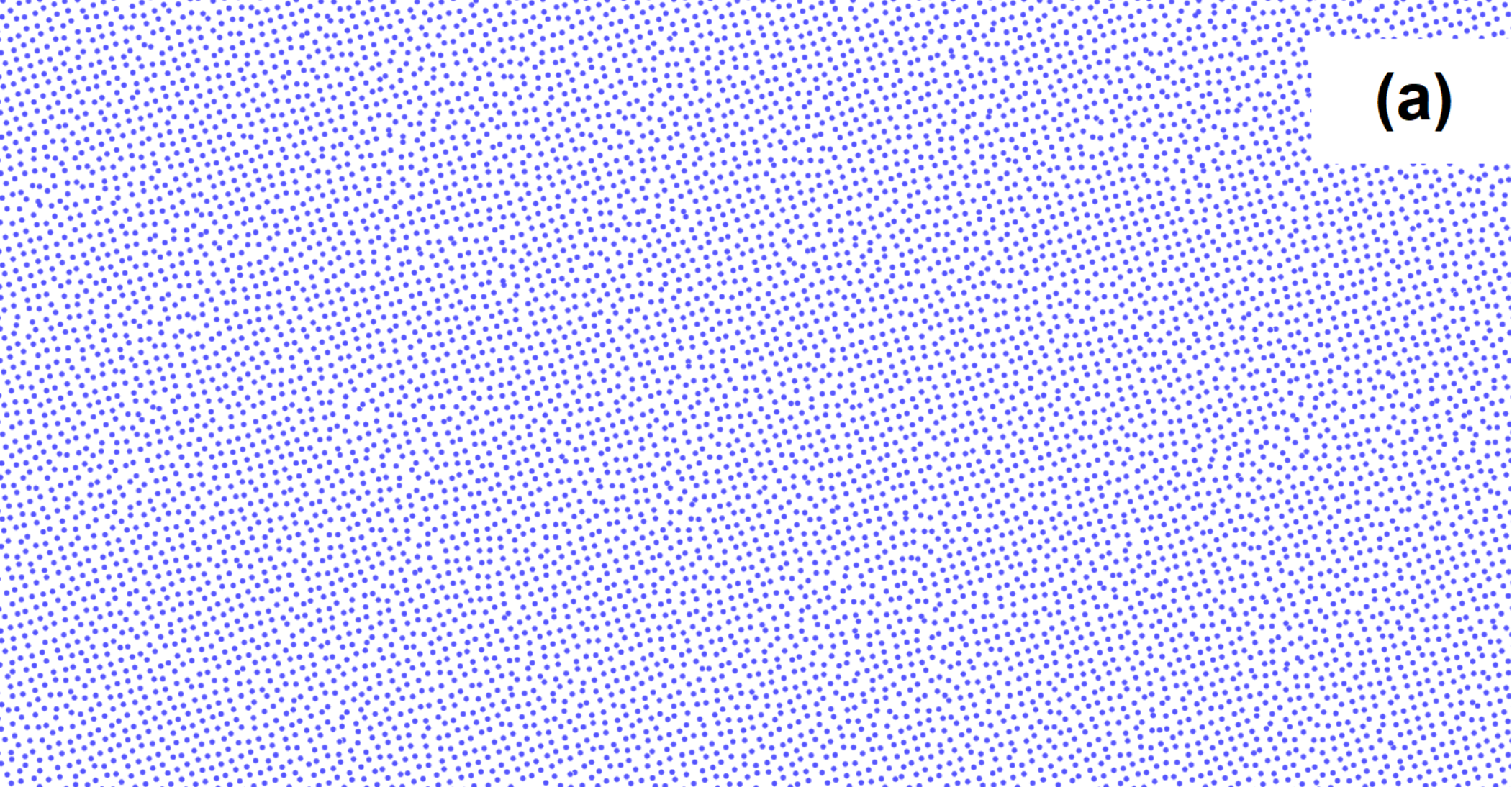}%

\includegraphics[width=8cm, height=6cm]{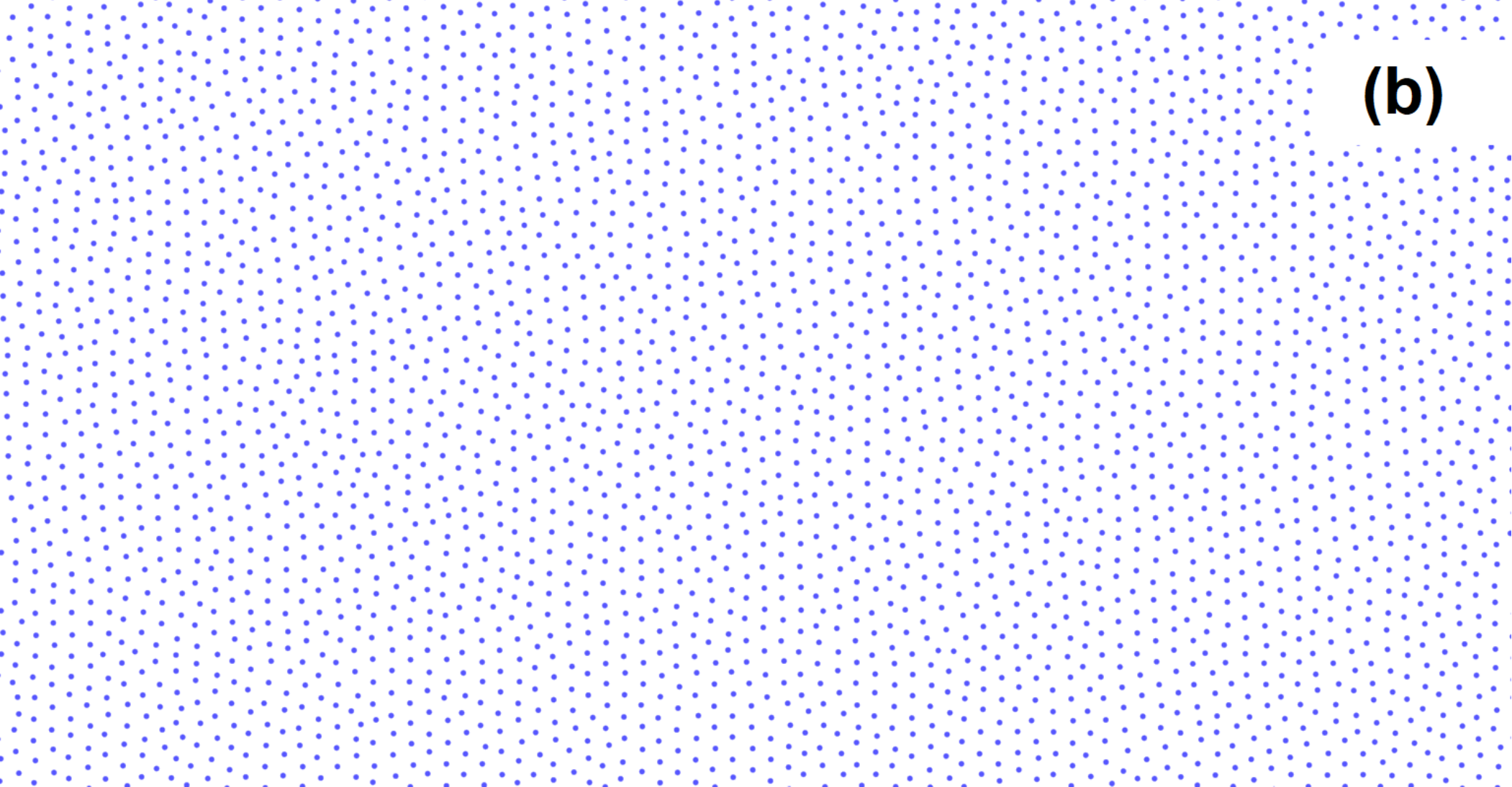}%

\includegraphics[width=8cm, height=8cm]{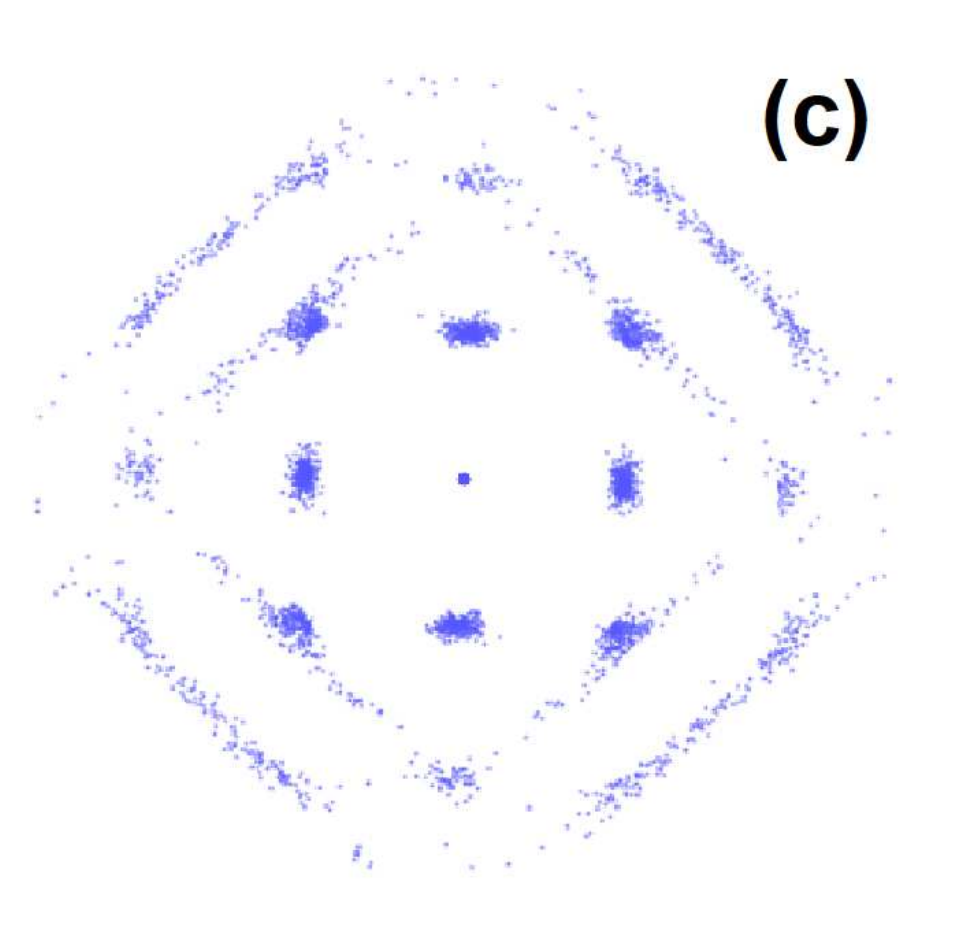}%

\caption{\label{r05} (a) A snapshot of the system at average density 0.5. (b) A snapshot of the first layer of the same system. (c) Diffraction pattern of the same system.}
\end{figure}

A snapshot of the system at $\rho=0.5$ is given in Fig. \ref{r05} (a). A snaphot of the first layer of the system is shown in the panel (b) of the same figure. The diffraction pattern at this density shows square symmetry (Fig. \ref{r05} (c)). At the same time both layers of the system demonstrate triangular lattice. Such a picture can appear if the layers are shifted with respect to each other. This result is consistent with the view that the structure can be considered as a cut of FCC (or HCP) lattice: the AB stacking gives a diffraction pattern with square symmetry.

\begin{figure}

\includegraphics[width=8cm, height=6cm]{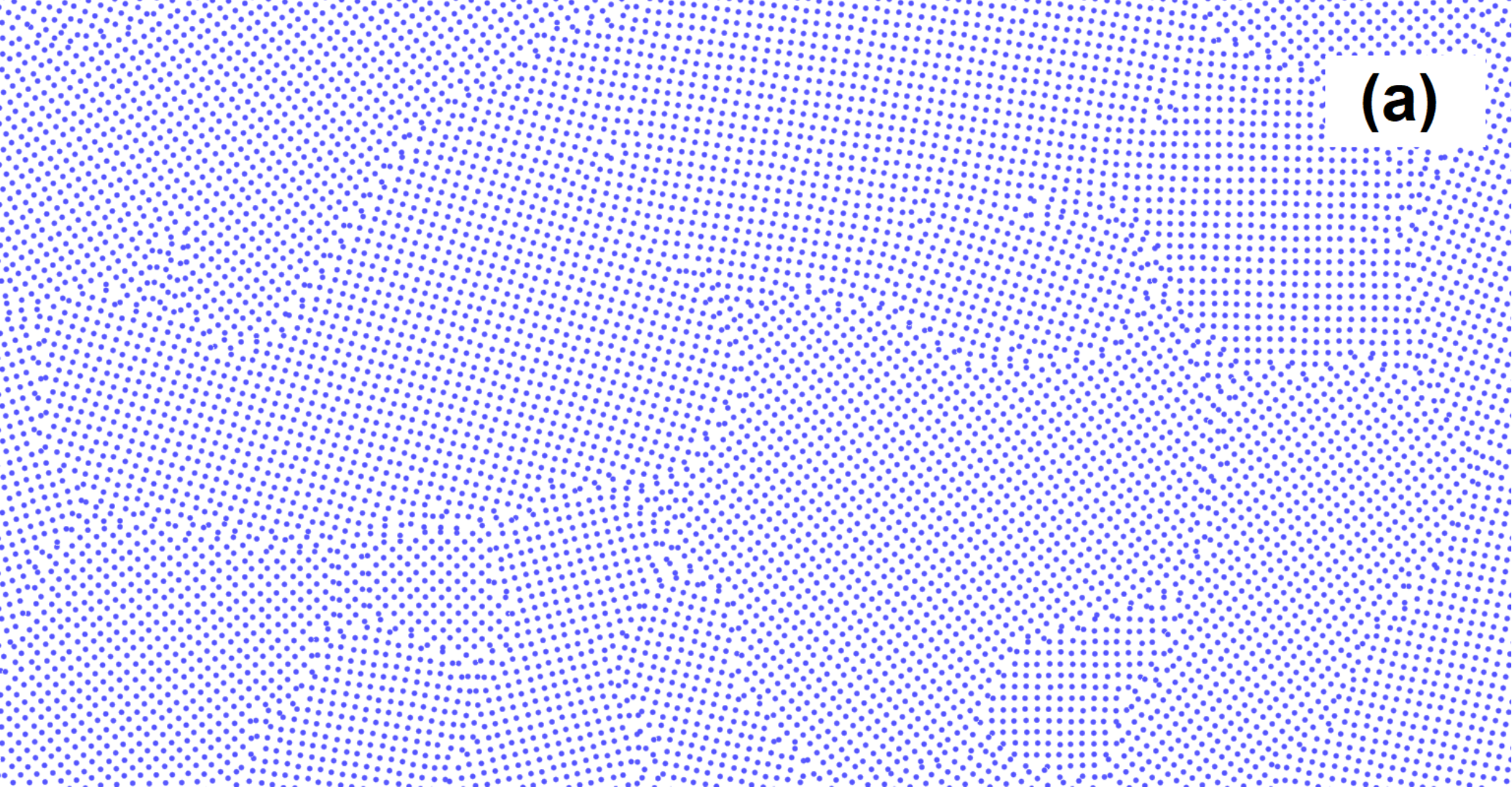}%

\includegraphics[width=8cm, height=6cm]{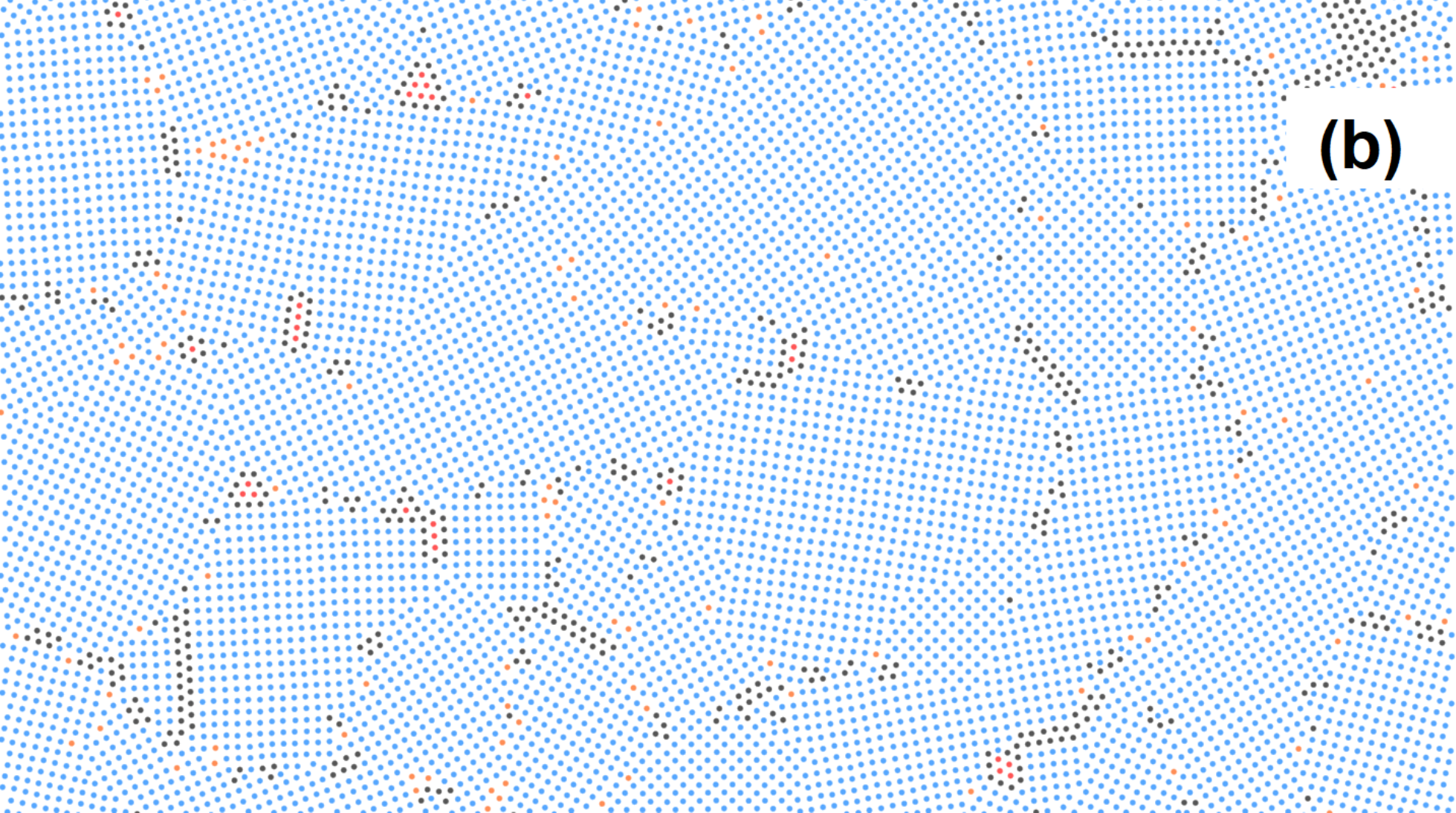}%

\includegraphics[width=8cm, height=8cm]{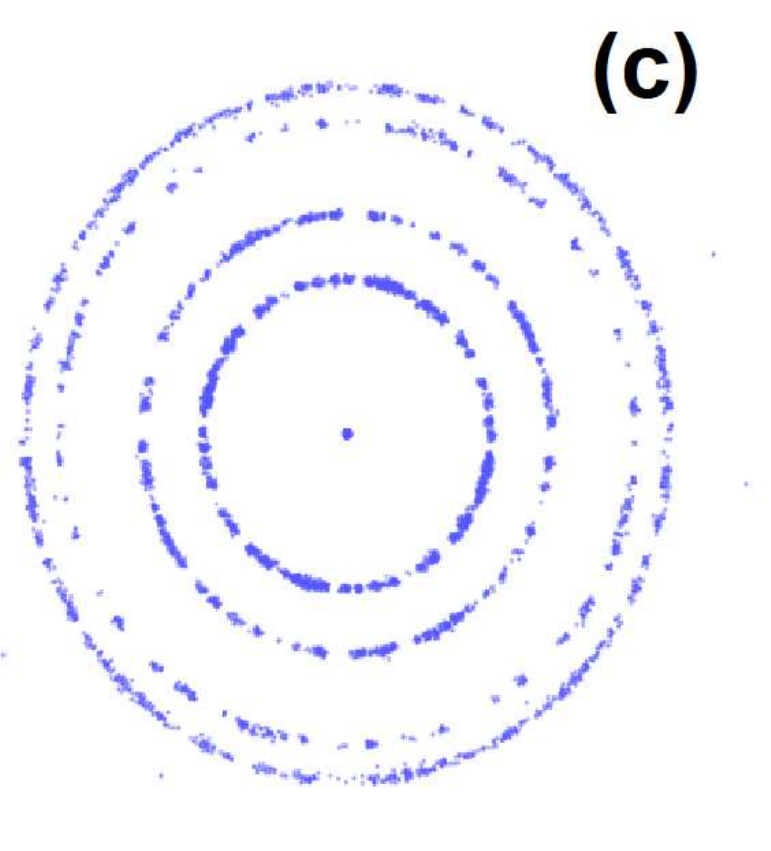}%

\caption{\label{r06} (a) A snapshot of the system at average density 0.6. (b) A snapshot of the first layer of the same system. Different colors correspond to different number of nearest neighbors of the particle and serve to make the grains easier to be seen. (c) Diffraction pattern of the same system.}
\end{figure}

At the density $\rho=0.6$ the system crystallized in two layers with square symmetry (Figs. \ref{r06} (a) and (b)). Both layers split in numerous grains which makes the diffraction pattern to be smeared (Fig. \ref{r06} (c)).

\subsection{Three layers}

\begin{figure}

\includegraphics[width=8cm, height=6cm]{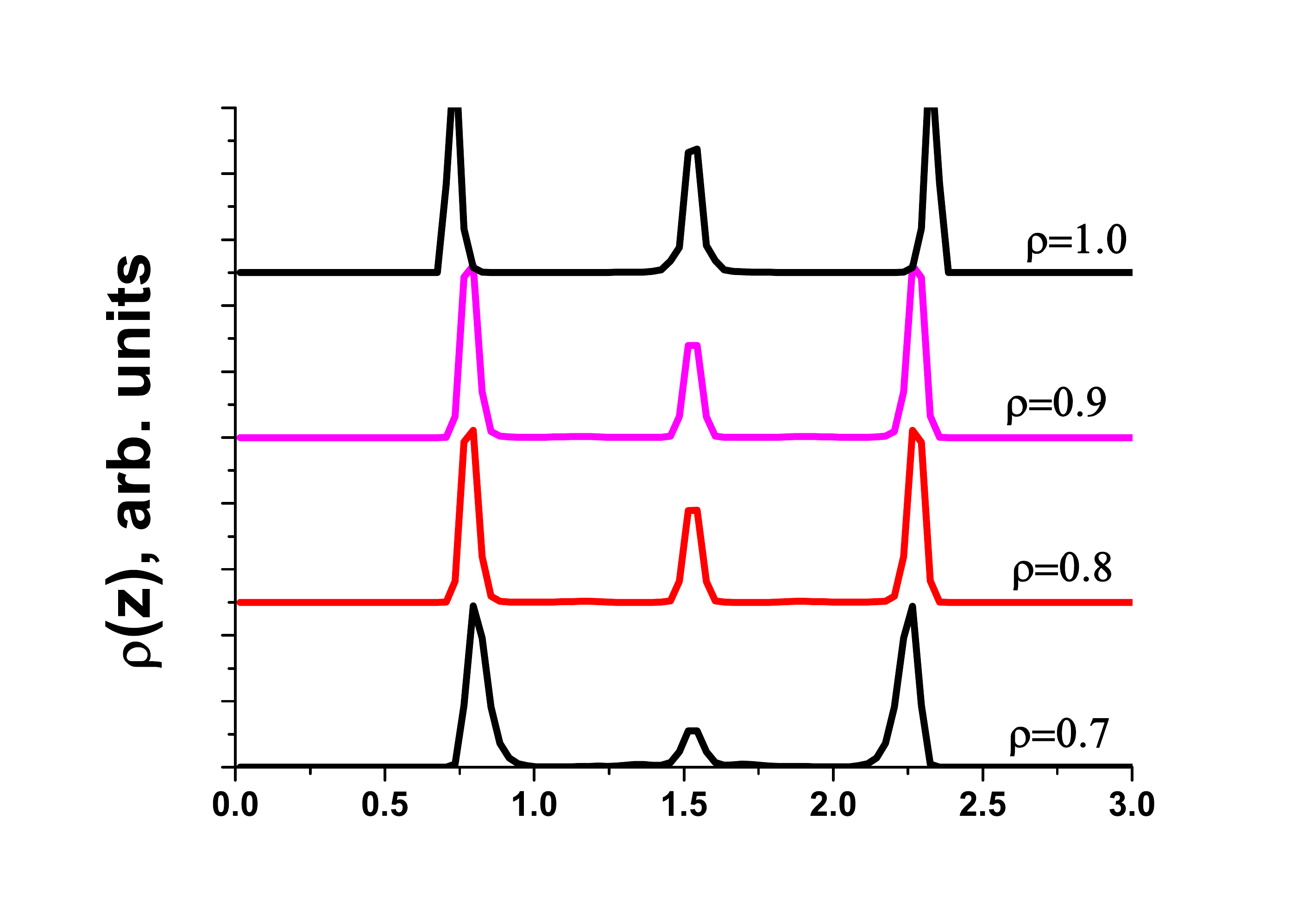}%

\includegraphics[width=8cm, height=6cm]{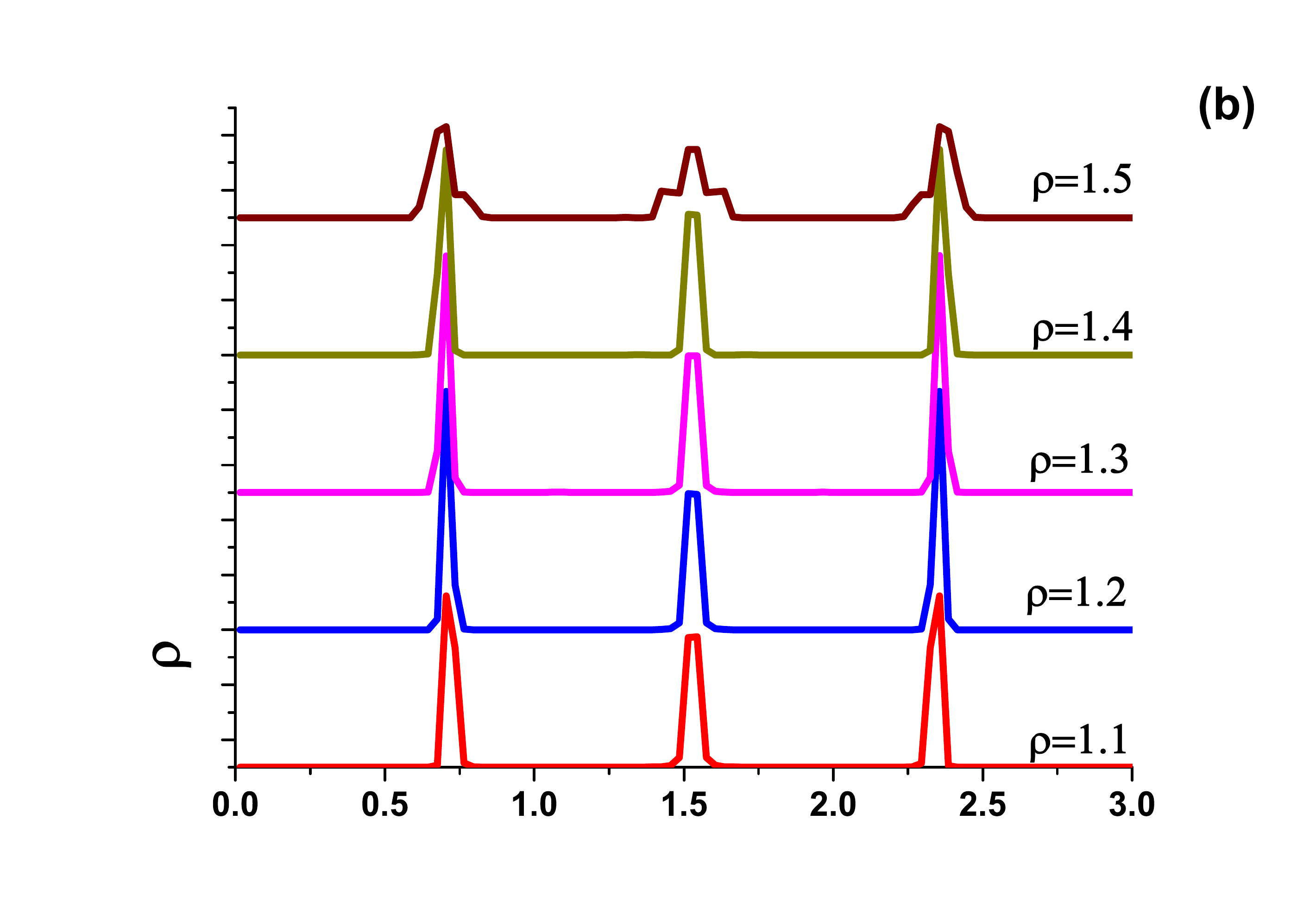}%

\caption{\label{z0715} (a) Distribution of density along z axis for the average densities from 0.7 to 1.0. (b) The distribution of the density along z axis for the average densities from 1.1 to 1.5 The curves are shifted with respect to each other to make the figure more clear.}
\end{figure}

Starting from the density $\rho=0.7$ the system splits into three layers. The distribution of the density along z axis is shown in Figs. \ref{z0715} (a) and (b).  Also the structure of the system experiences dramatic changes.

\begin{figure}

\includegraphics[width=8cm, height=6cm]{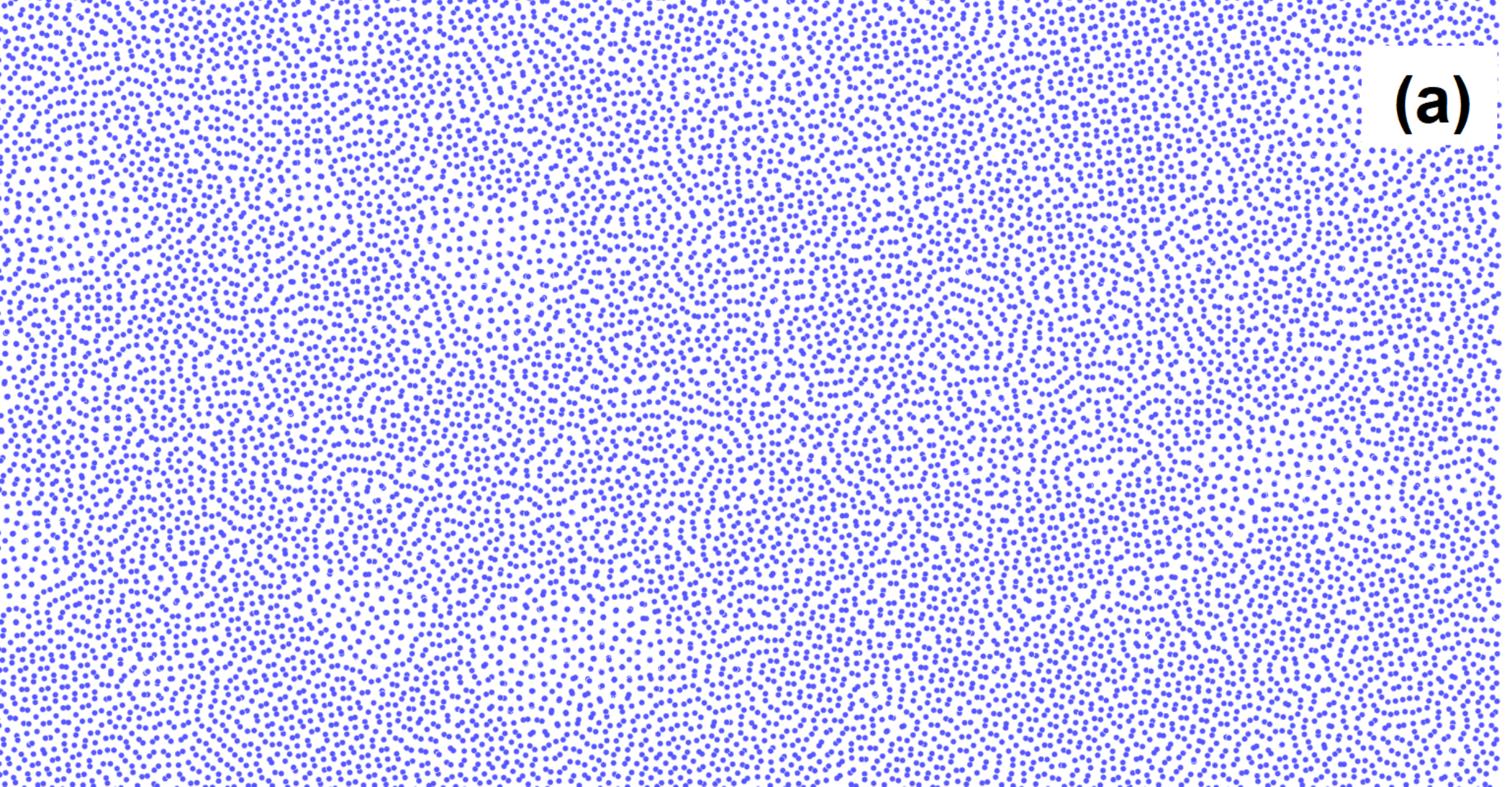}%

\includegraphics[width=8cm, height=6cm]{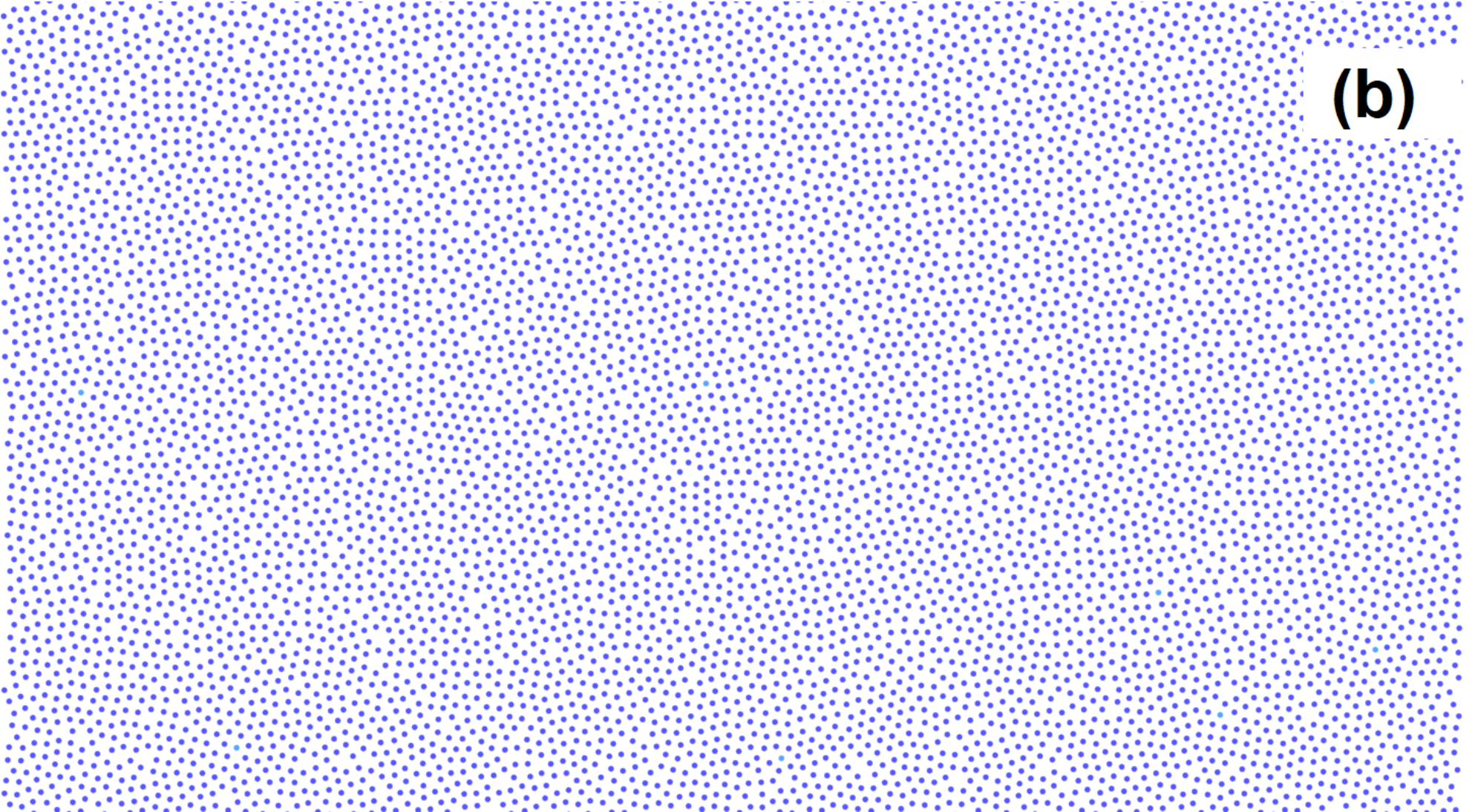}%

\includegraphics[width=8cm, height=6cm]{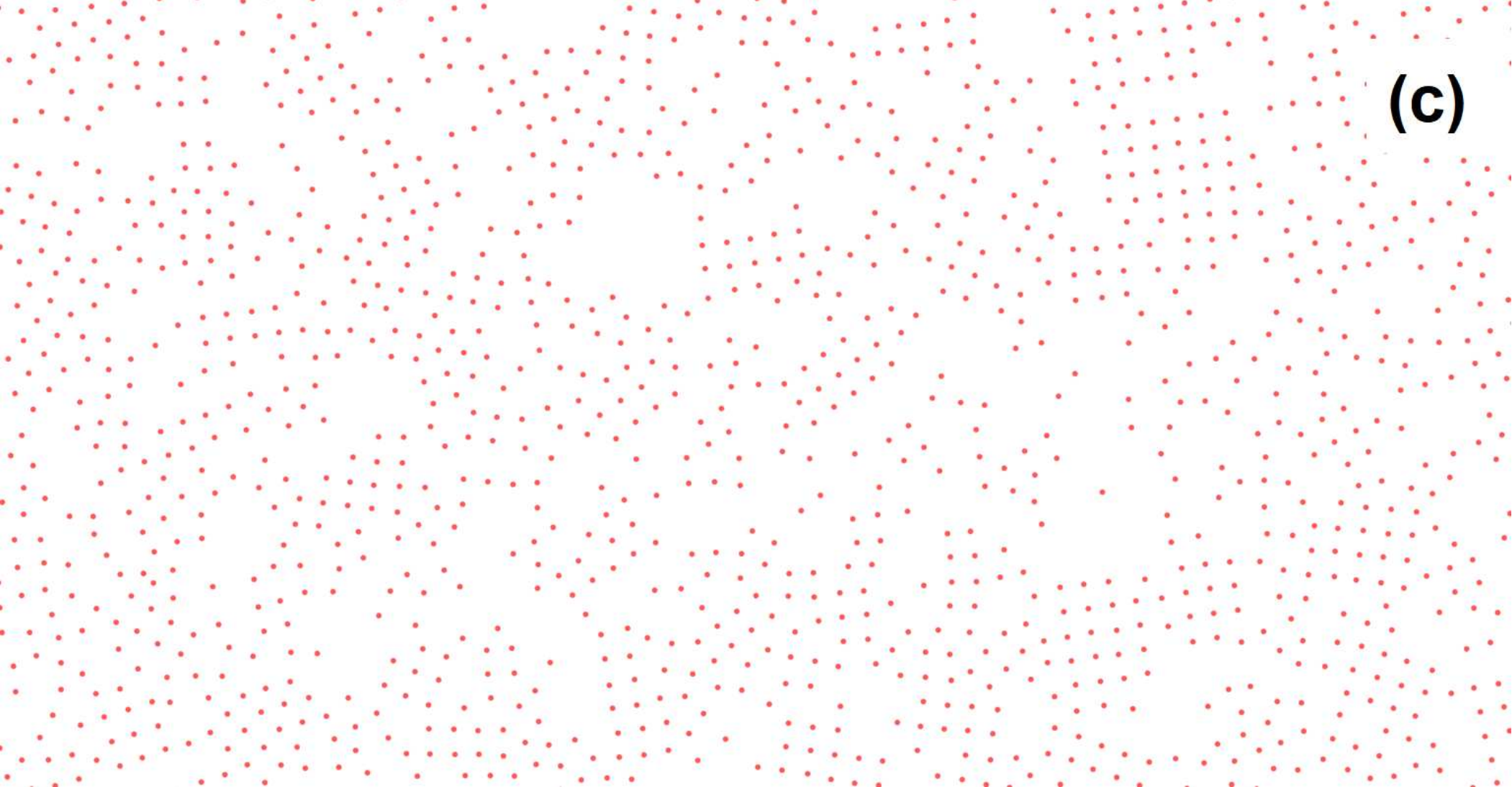}%

\caption{\label{r07} (a) A snapshot of the system at average density 0.7. (b) A snapshot of the outer (the first or the third) layer of the same system. (c) A snapshot of the inner (the second) layer of the same system.}
\end{figure}

Fig. \ref{r07} (a) shows a snapshot of the the system at $\rho=0.7$. One can see some elemenst of the structure which look like Moire patterns. In previous publications Moire patterns were observed in colloidal suspensions confined between two glass plates \cite{oguz1,moire}. It was discovered that such patterns correspond a structure which consists of several triangular layers which, however, are rotated with respect to each other. This is, however, not the case of the present study. First, we observe that the number of particles in the inner layer is much smaller than in the outer ones (about 4500 in the former and 13750 in the later). Fig. \ref{r07} (b) and (c) show snapshots of the outer (the first and the third) and the inner (the second) layers, respectively. In the case of the outer layers no visible structure is observed, while in the case of the inner one clusters of square crystal are visible. However, the structure is very defected. In particular, it is not completely filled, since numerous holes are observed in the inner layer.

\begin{figure}

\includegraphics[width=8cm, height=6cm]{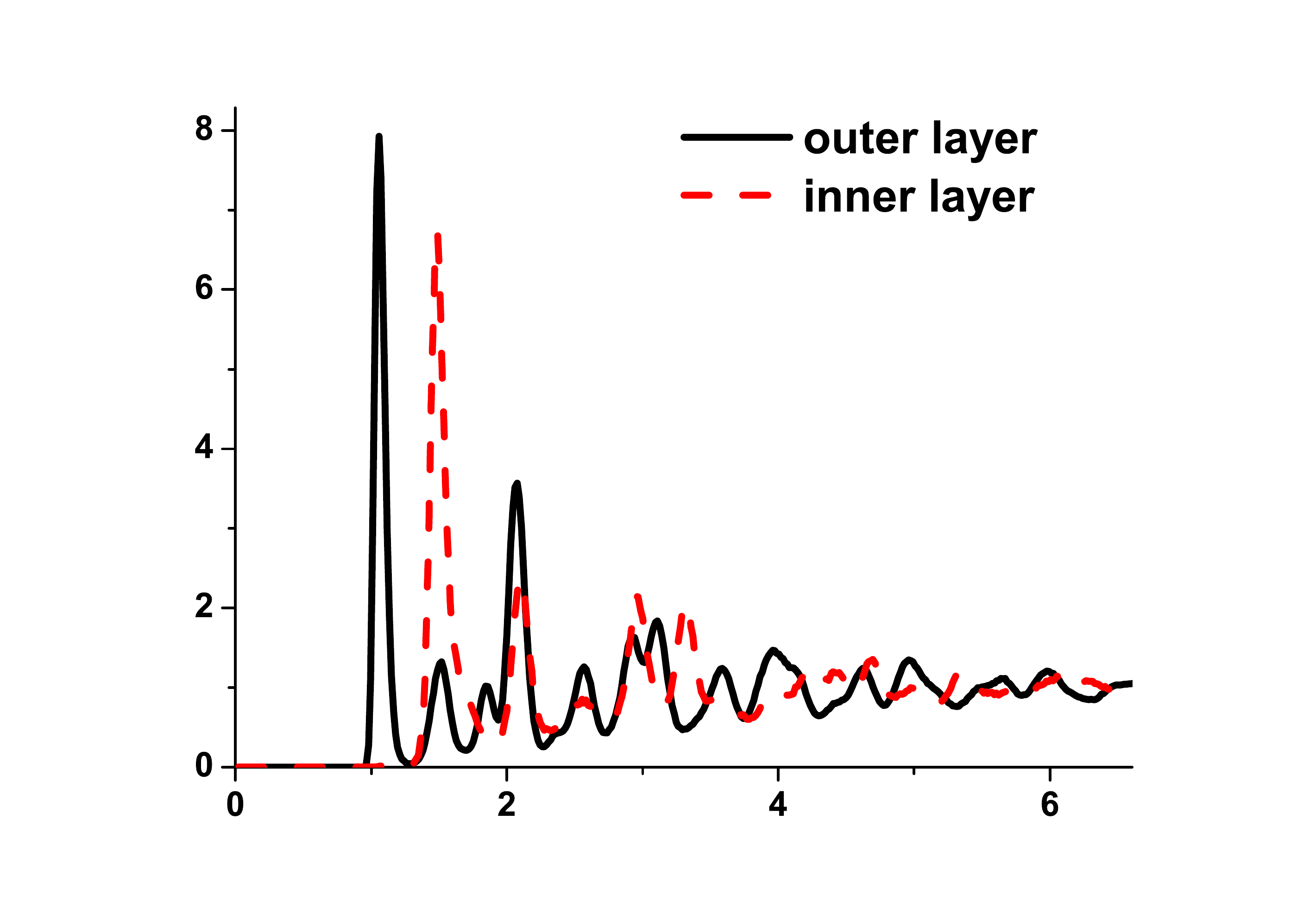}%

\caption{\label{rdf07} Two-dimensional radial distribution functions of the outer and inner layers of the system at average density $\rho=0.7$.}
\end{figure}

Fig. \ref{rdf07} shows RDFs of the outer and inner layers. In both cases the RDFs demonstrate several peaks, which means that there is strong local ordering in the system. However, after several peaks the RDFs smear and go to the unity, i.e., no long-range order is observed. Importantly, the first peak of RDF of the outer layers is located at $r_1=1.05$, while the location of the first peak of the RDF of the inner layer coincides with the one of the second peak of the outer one $r_2=1.49$. This effect can be related to the presence of the holes in the inner layer of the particles.

\begin{figure}

\includegraphics[width=8cm, height=6cm]{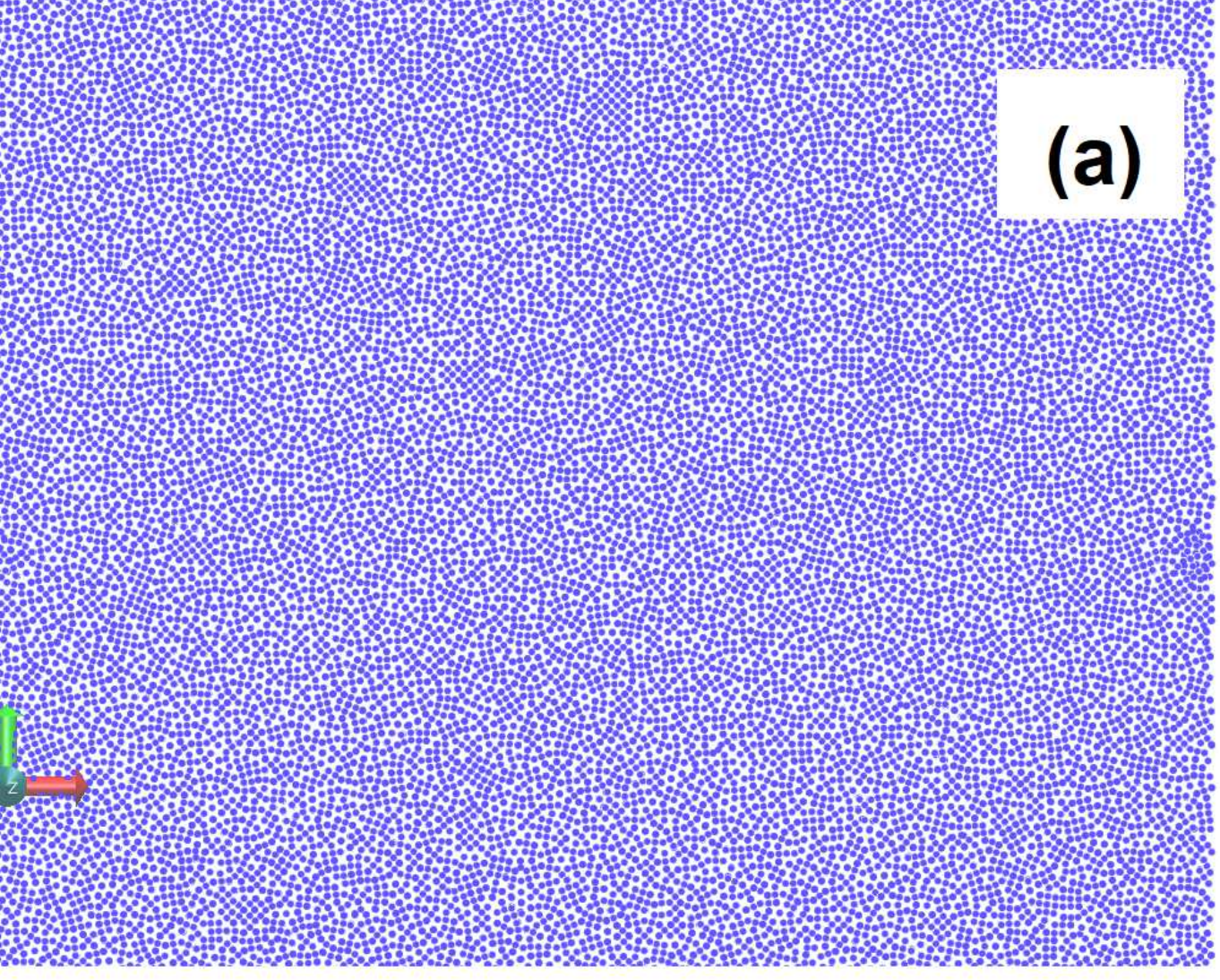}%

\includegraphics[width=8cm, height=6cm]{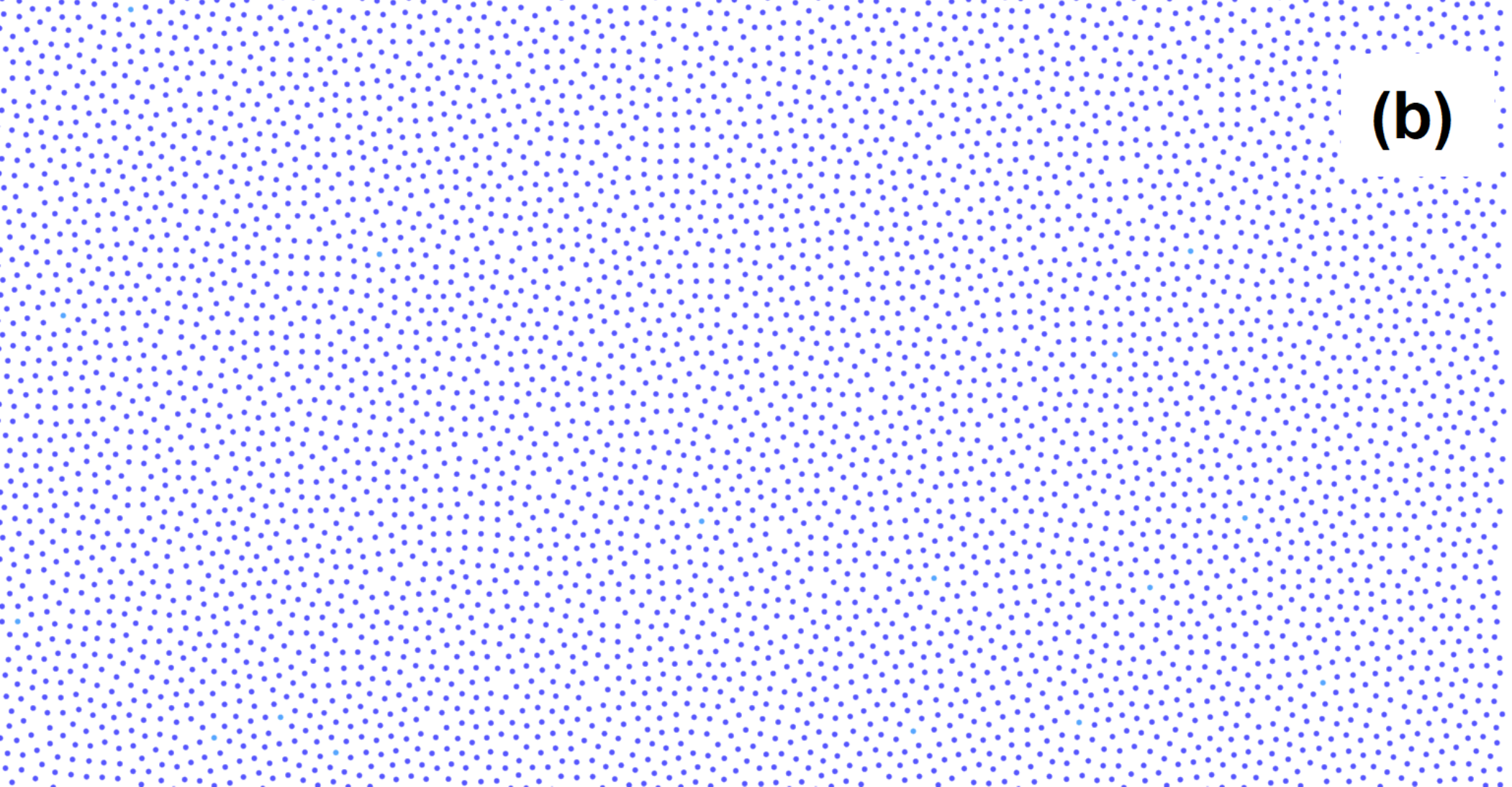}%

\includegraphics[width=8cm, height=6cm]{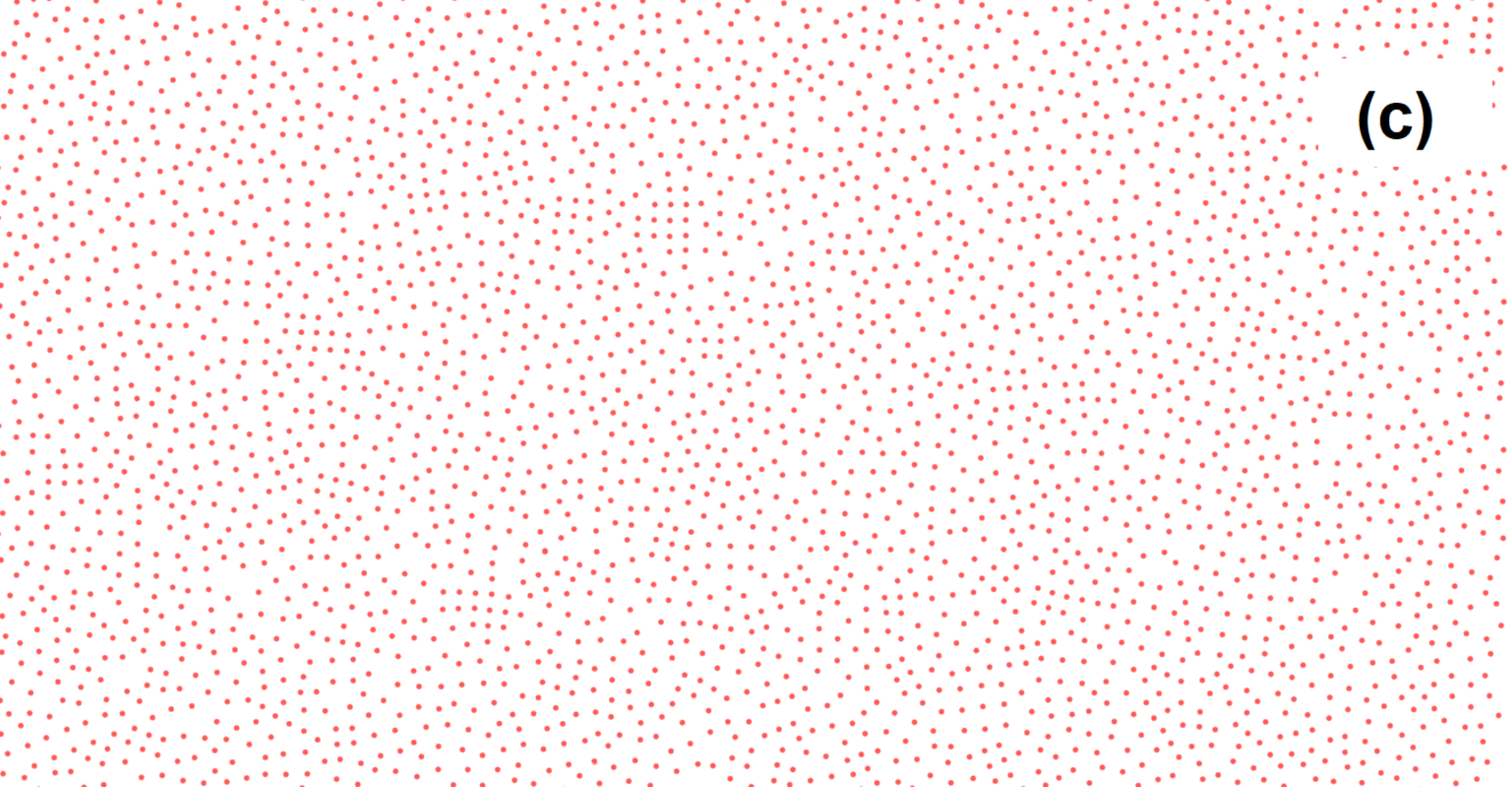}%

\caption{\label{r08} (a) A snapshot of the system at average density 0.8. (b) A snapshot of the outer layer of the same system. (c) A snapshot of the inner layer of the same system.}
\end{figure}

Fig. \ref{r08} (a) shows a snapshot of the system at $\rho=0.8$. It does not demonstrate apparent ordering. No ordering is observed also within the layers (panels (b) and (c)).

\begin{figure}

\includegraphics[width=8cm, height=6cm]{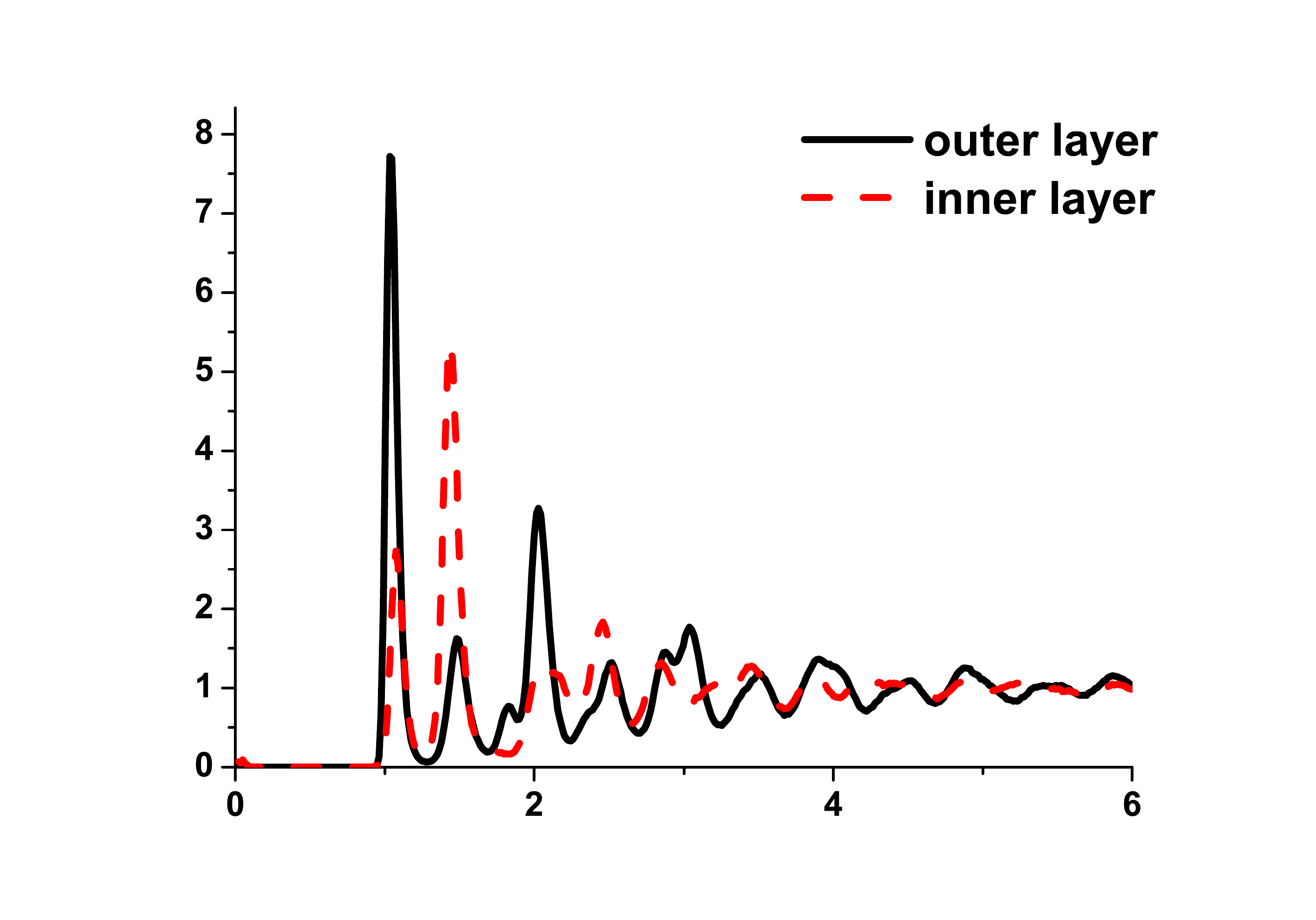}%

\caption{\label{rdf08} Two-dimensional radial distribution functions of the outer and inner layers of the system at average density $\rho=0.8$.}
\end{figure}

The number of particles in the inner layer is larger than in the previous case. At $\rho=0.8$ the inner layers contains 7270 particles in average. It results in the changes in the RDF of the inner layer which is qualitatively similar to the changes of the RDF in purely 2d or purely 3d core-softened systems: the particles go on the shoulder of the potential, which results in appearance of small first peak at $r_1=1.075$ (Fig. \ref{rdf08}). In both 2d and 3d systems this leads to the anomalous melting of the crystal with negative slope of the melting line. In the present study we also observe that the structure of the second (inner) layer becomes less pronounced than in the case of density 0.7.

\begin{figure}

\includegraphics[width=8cm, height=6cm]{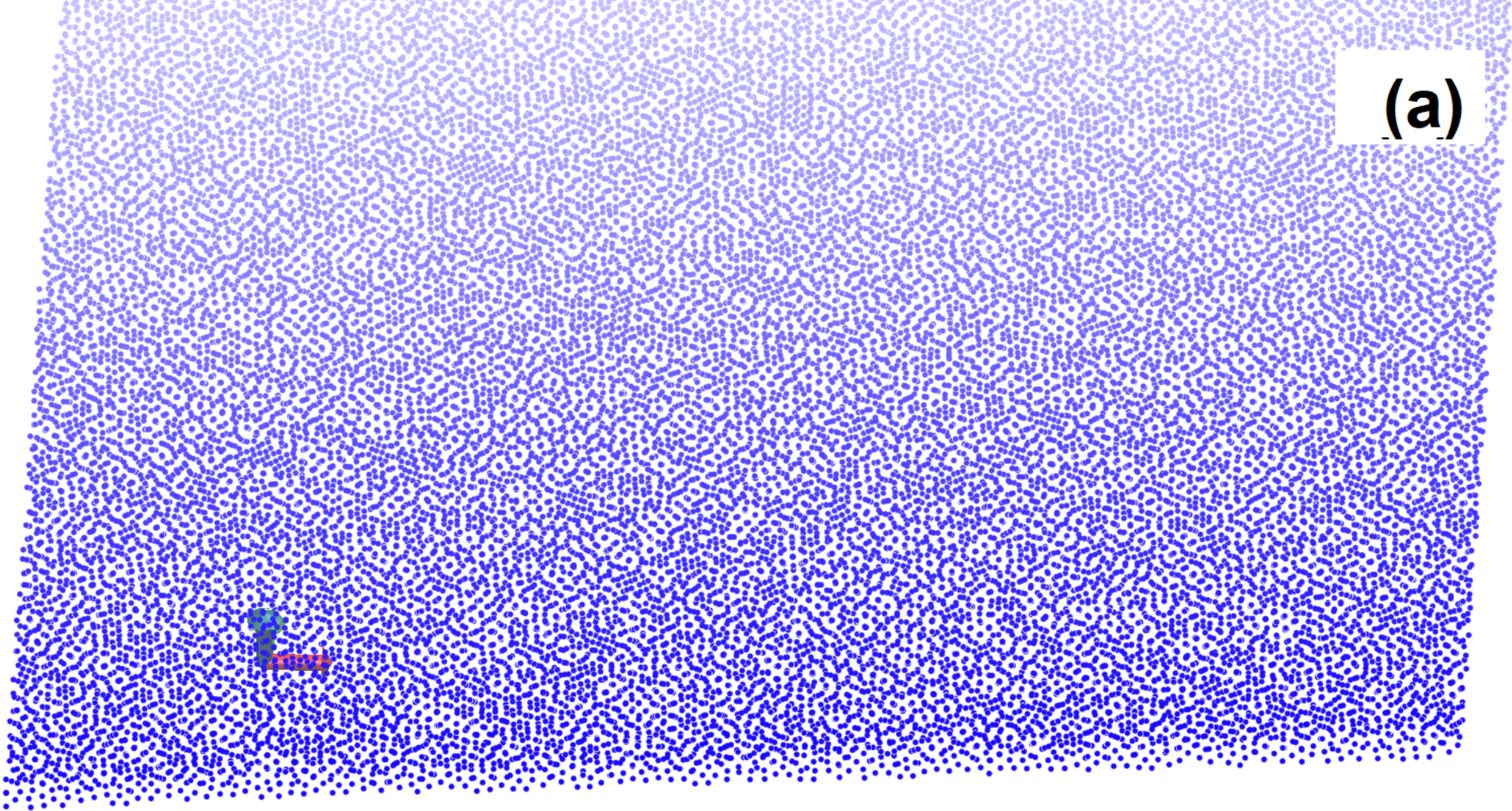}%

\includegraphics[width=8cm, height=8cm]{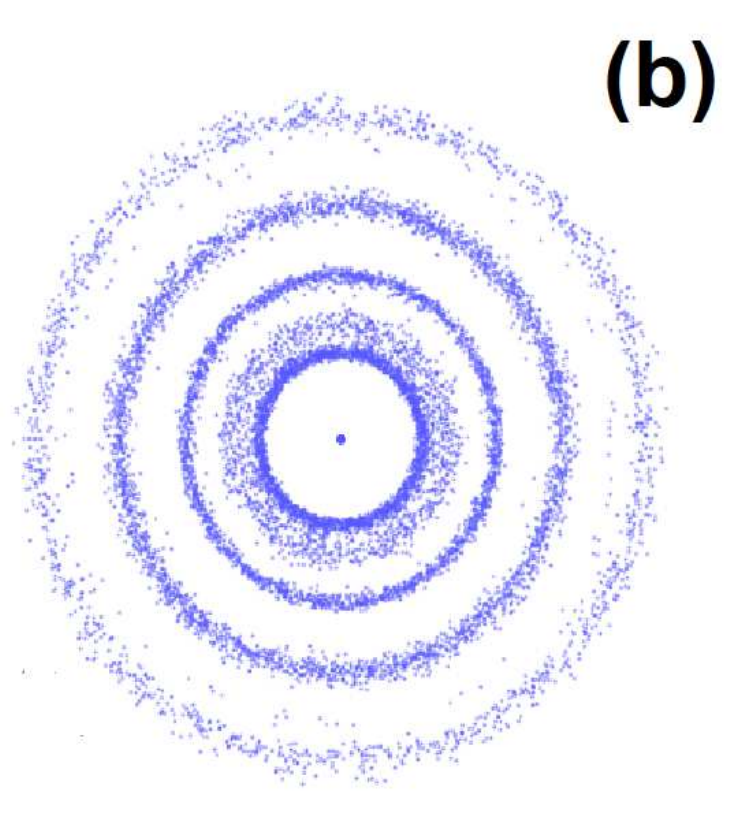}%

\includegraphics[width=8cm, height=8cm]{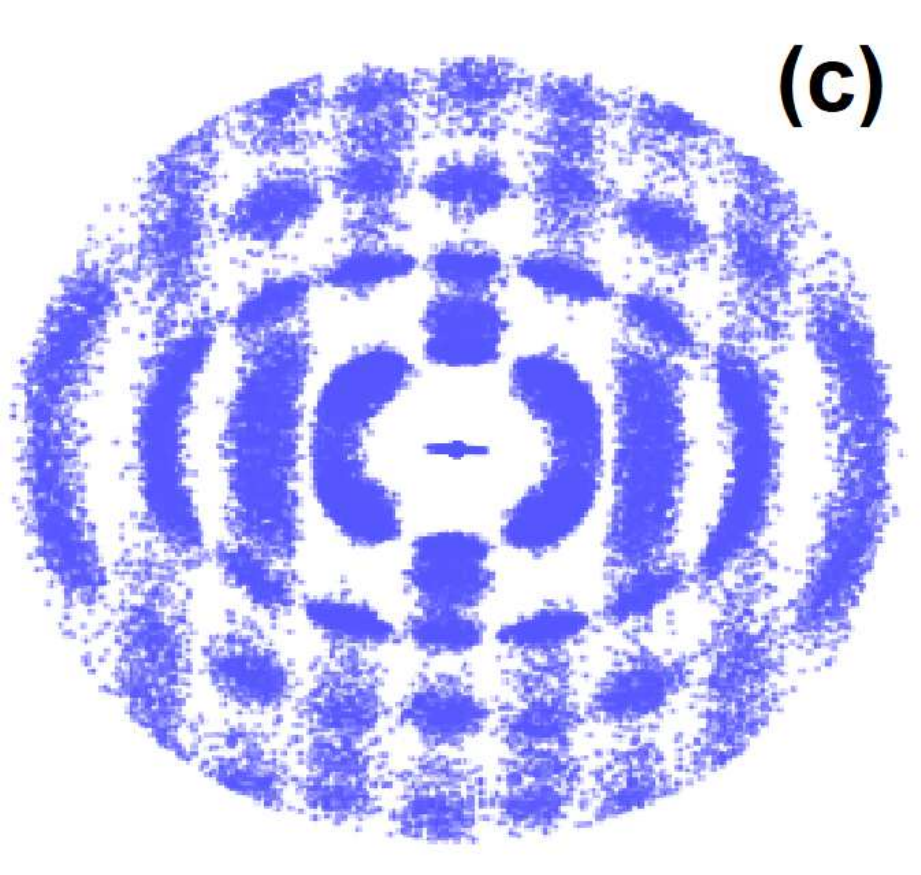}%

\caption{\label{r08-turned} (a) A snapshot of the system at average density 0.8. (b) Diffraction pattern of the system at $\rho=0.8$ without rotation. (c) Diffraction pattern of the system at $\rho=0.8$ after rotation to 45 degrees.}
\end{figure}

In our previous study we demonstrated that in the case of small pores the stcurture of the LJ system can be considered as rotation of HCP or FCC one with respect to the walls of the pore. Following this idea we study the structure of the present system at $\rho=0.8$ from different angles. Fig. \ref{r08-turned} (a) shows a snapshot of the system rotated by 45 degrees around x-axis. One can see some elements characteristic to quasicrystalline phases, similar to the ones observed for the density 0.7. Fig. \ref{r08-turned} (b) and (c) demonstrate the diffraction pattern before the rotation and after the rotation. While the former does not demonstrate any sturcture, twelve peaks are observed in the later. This is consistent with the existence of the dodecagonal quasicrystal in purely 2d RSS.

From this observation we may assume that the system forms a quasicrystal which is rotated with respect to the planes of the walls.

The behavior of the system at density 0.9 is qualitatively analogous to the one of the density 0.8.

\begin{figure}

\includegraphics[width=8cm, height=6cm]{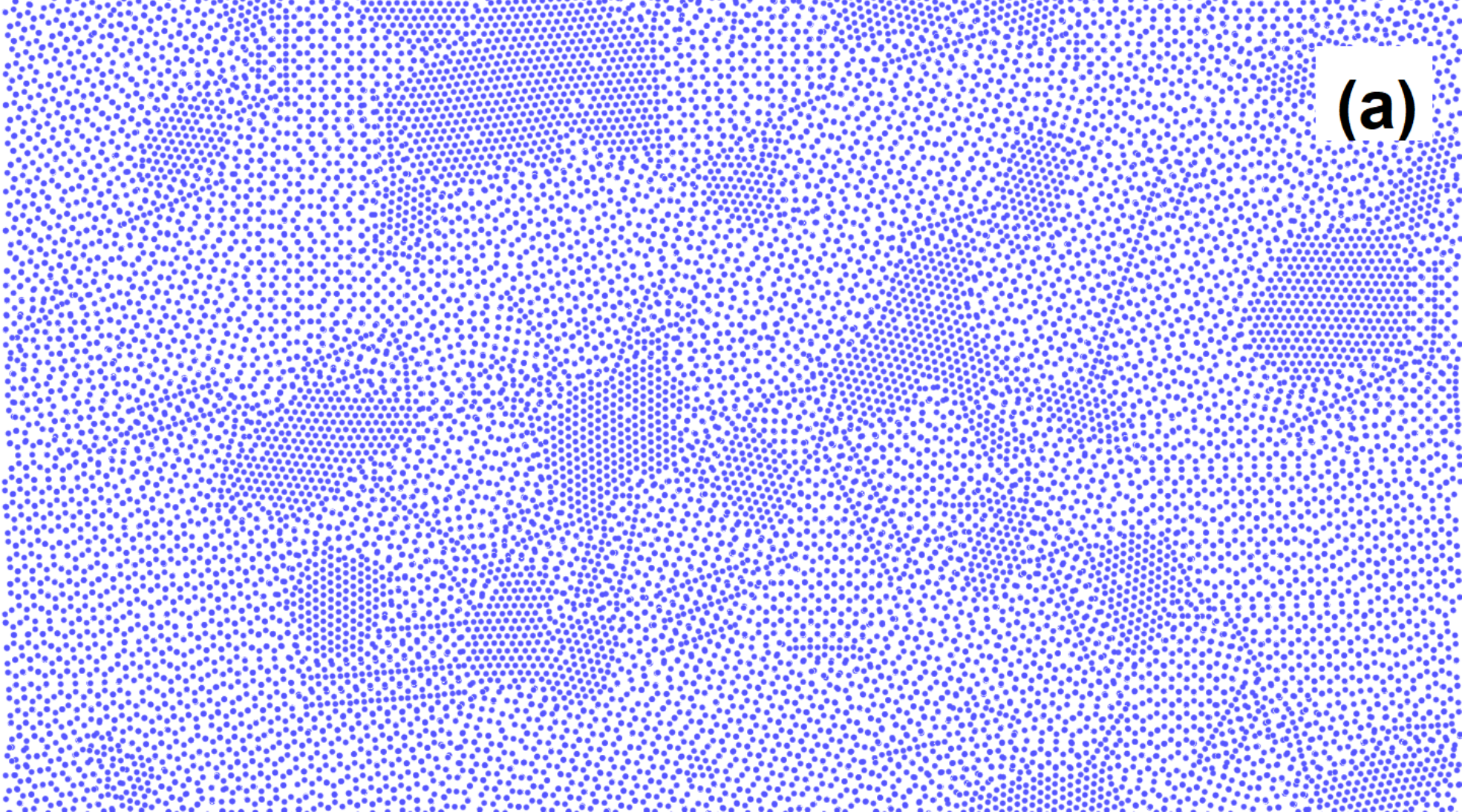}%

\includegraphics[width=8cm, height=8cm]{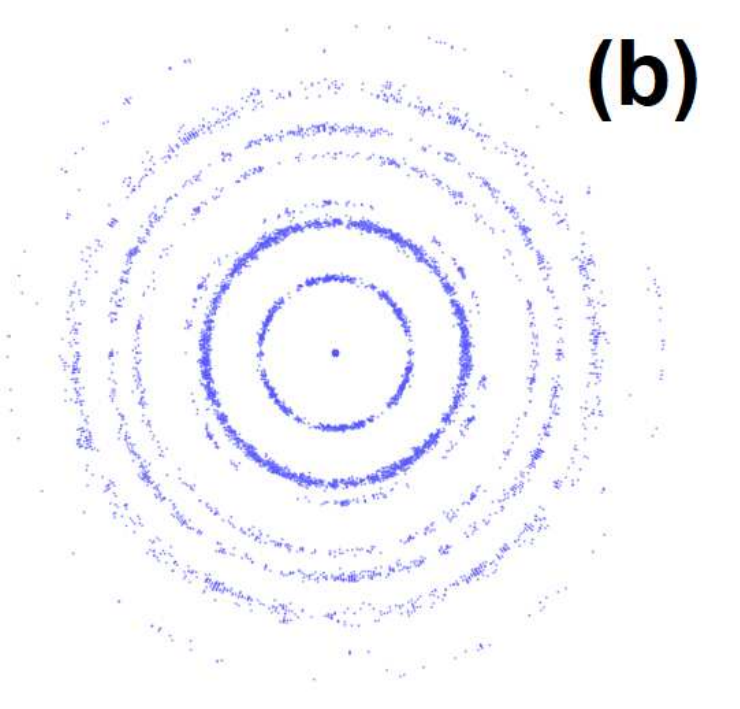}%

\includegraphics[width=8cm, height=6cm]{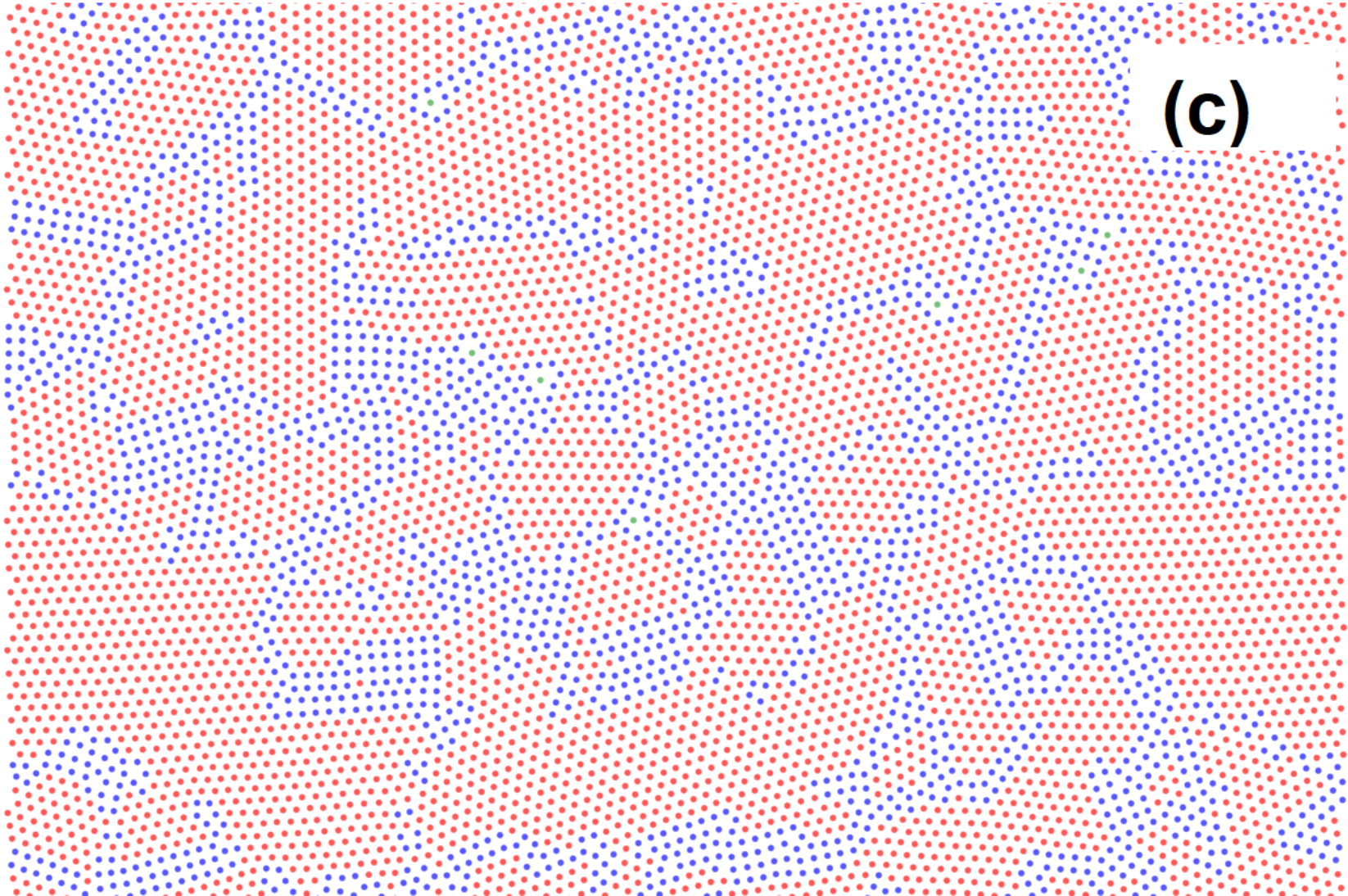}%

\includegraphics[width=8cm, height=6cm]{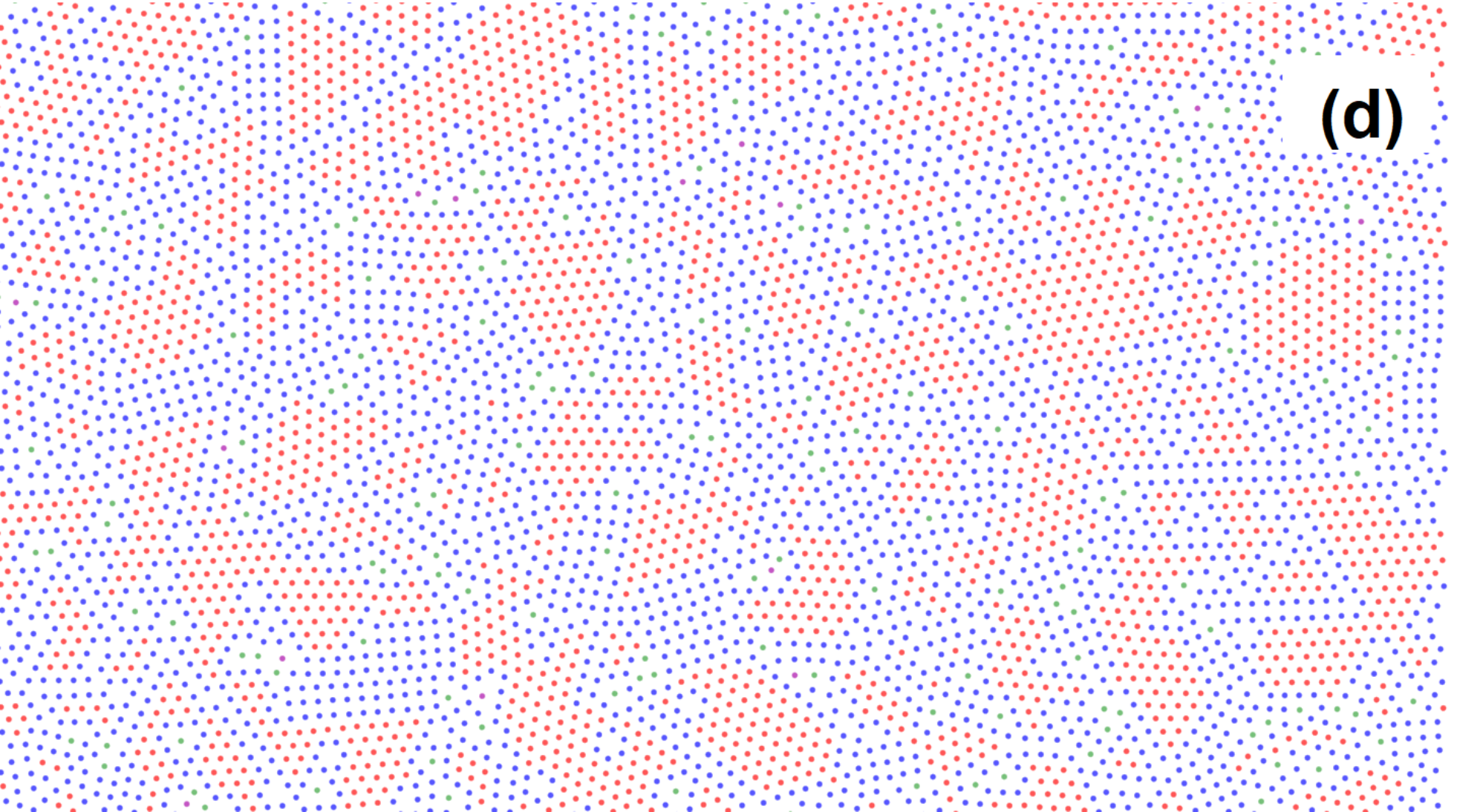}%

\caption{\label{r1} (a) A snapshot of the system at average density 1.0. (b) Diffraction pattern of the same system (c) A snapshot of the outer layer of the same system. (d) A snapshot of the inner layer of the same system.}
\end{figure}

\begin{figure}

\includegraphics[width=8cm, height=6cm]{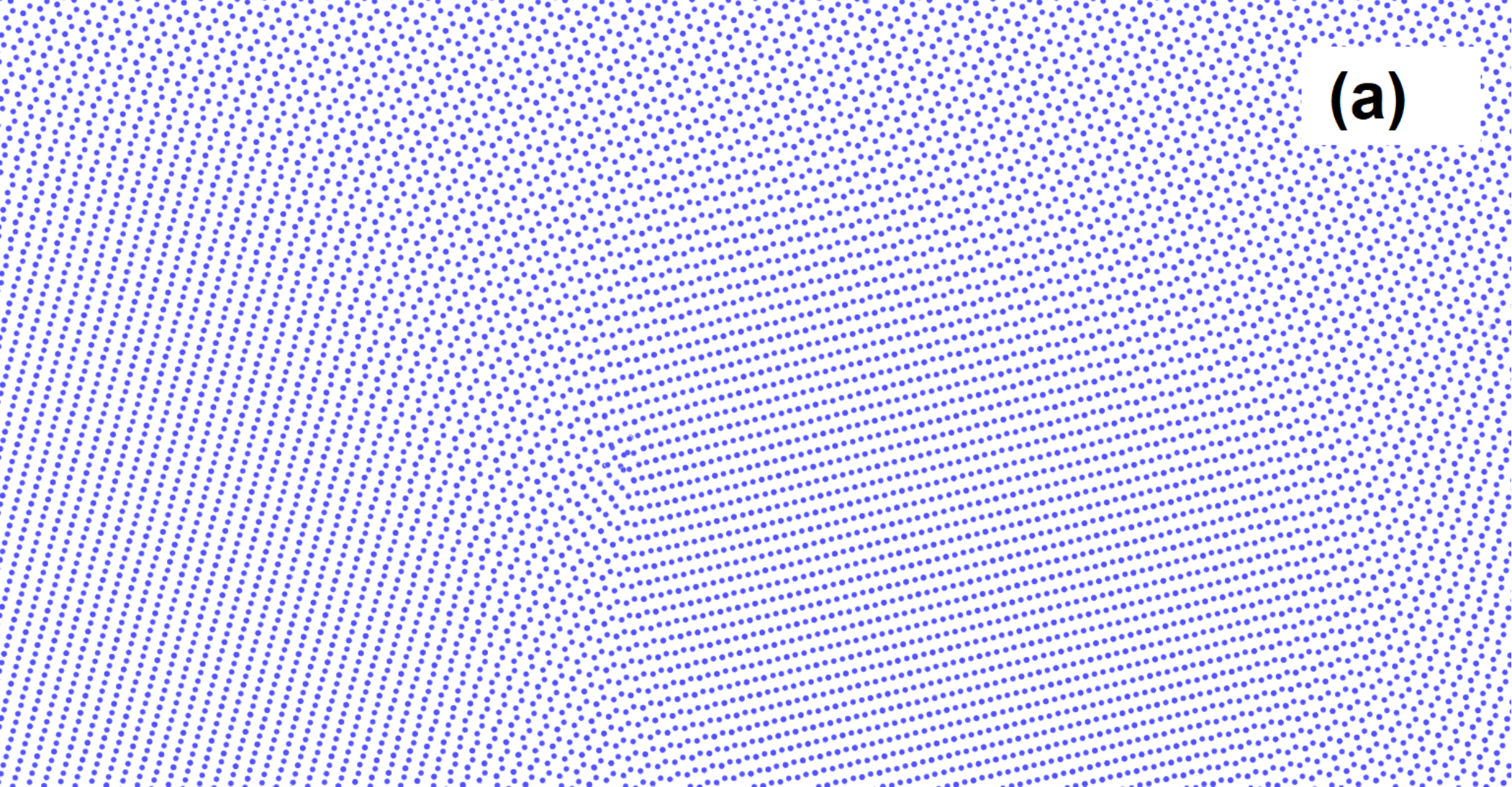}%

\includegraphics[width=8cm, height=8cm]{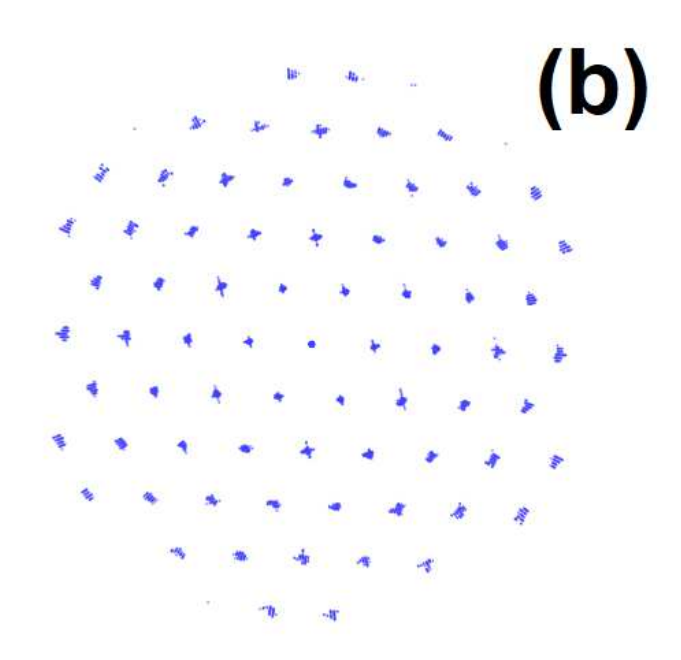}%

\includegraphics[width=8cm, height=6cm]{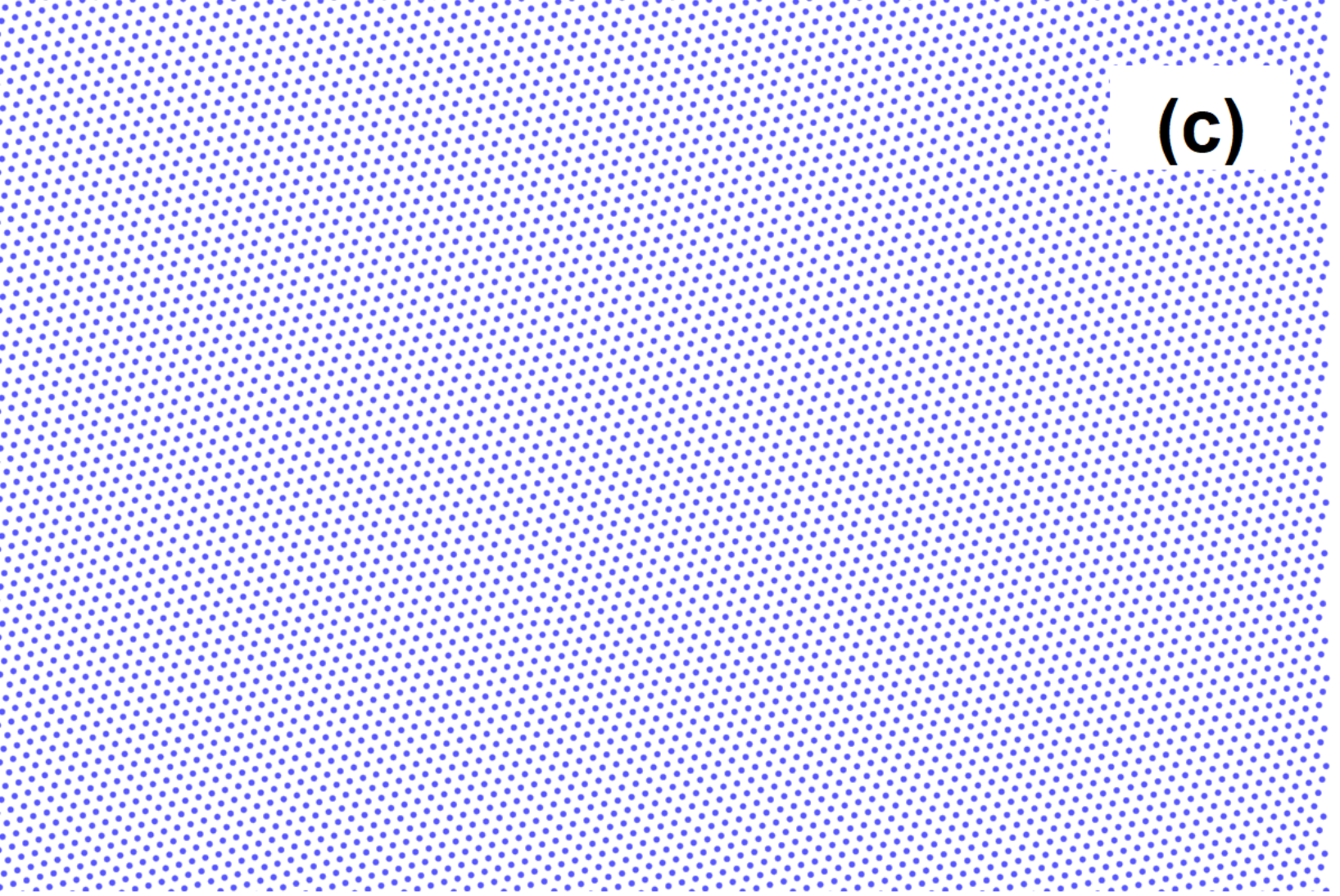}%

\includegraphics[width=8cm, height=6cm]{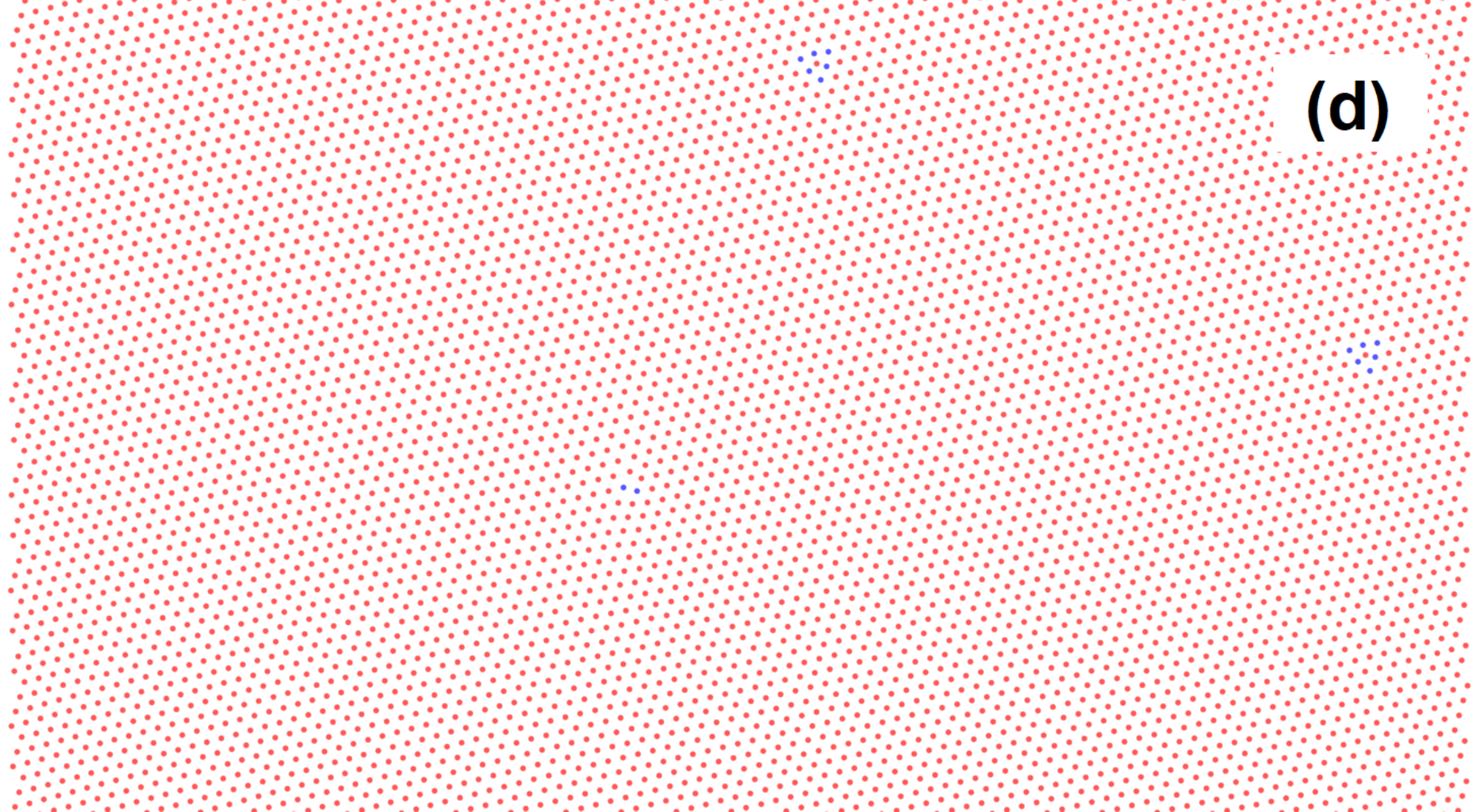}%

\caption{\label{r15} (a) A snapshot of the system at average density 1.5. (b) Diffraction pattern of the same system (c) A snapshot of the outer layer of the same system. (d) A snapshot of the inner layer of the same system.}
\end{figure}

Starting from the density $\rho=1.0$ we again observe that the system crystallizes into numerous grains of FCC or HCP structure. Fig. \ref{r1} and \ref{r15} give examples of snapshots and diffraction patterns of systems at $\rho=1.0$ (the former) and $\rho=1.5$ (the later). One can see that at the density $\rho=1.0$ the grains are small which leads to the smeared diffraction patterm, while at $\rho=1.5$ the grains are large which results in clear FCC- or HCP-like diffraction pattern.

Summurizing the results, we observe the following sequence of phases: HCP (or FCC) with triangular structure of the layers (densities from 0.3 to 0.5), HCP (or FCC) with square structure of the layers ($\rho=0.6$), quasicrystal (densities 0.7 - 0.9), HCP with triangular structure of the layers ($\rho>=1.0$). This sequence reminds the one of the purely 2d RSS: triangular crystal - square crystal - dodecagonal quasicrystal - triangular crystal. Therefore, the phase diagram of the quasi-2d RSS is related to the one of the purely 2d, but some phases are smeared.

Another point of interest is that increasing of the density of the systems with two layers leads to change of the in-layer structure from the triangular one to the square one, which is in contrast with simple systems, where the structures with triangular in-layer crystals correspond to larger density than the ones with the square one. This effect can be related to the general features of the core-softened systems. At low density these systems crystallize at the density with lattice constant corresponding to the width of the repulsive shoulder. When the density is increased, the particles penetrate on the shoulder which leads to melting of the systems. Further densification leads to formation of low-coordinated structures, such as the ones with square in-layer ordering and quasicrystals in the present case.

\section{Conclusions}

This work discusses the structure of RSS in strong confinement (two of three layers of the particles in a slit pore). We observe that the sequence of phases is more complex than in the case of simple systems, such as Hard Spheres or LJ one. The sequence of the phases is similar to the one of the purely 2d system, but the structures are smeared. Also, the sequence of phases is not the same as in the simple systems (two layers with triangular structure correspond to lower densities than two layers with square structure).

These results show that systems with complex interaction demonstrate more complicated behavior in confinement than the ones with simple interactions. This result is  important for investigation of the behavior of confined systems with complex interactions, for instance, metals and water.

This work was carried out using computing resources of the federal
collective usage center "Complex for simulation and data
processing for mega-science facilities" at NRC "Kurchatov
Institute", http://ckp.nrcki.ru, and supercomputers at Joint
Supercomputer Center of the Russian Academy of Sciences (JSCC
RAS). The work was supported by the Russian Science
Foundation (Grants No 19-12-00092).

\end{document}